\documentclass[a4paper,fleqn,usenatbib]{mnras}

\usepackage{graphicx}
\usepackage{amsmath}
\usepackage{amsfonts}
\usepackage{amssymb}
\usepackage{multirow}
\usepackage{times}
\usepackage[T1]{fontenc}
\usepackage{aecompl}
\usepackage[normalem]{ulem}
\usepackage{longtable}
\usepackage{multicol}
\usepackage{blkarray}
\usepackage{txfonts}
\usepackage{color}

\newcommand{\tsne}{{\tt t-SNE}}	
\newcommand{\tab}[1]{Table~\ref{#1}}
\newcommand{\fig}[1]{Figure~\ref{#1}}
\newcommand{\sect}[1]{Sect.~\ref{#1}}

\title[Parallaxes of RAVE-TGAS twin stars]{Climbing the cosmic ladder with stellar twins in RAVE with Gaia}

\author[P. Jofr\'e et al.]{
P. Jofr\'e$^{1,2}$\thanks{E-mail: pjofre@ast.cam.ac.uk}, 
G. Traven$^{3}$, K. Hawkins$^{4}$, G. Gilmore$^{1}$, J. L. Sanders$^{1}$, T. M\"adler$^{1}$,  
\newauthor M. Steinmetz$^{5}$, A. Kunder$^{5}$, G. Kordopatis$^{6}$, P. McMillan$^{7}$, O. Bienaym\'e$^{8}$, 
\newauthor  J. Bland-Hawthorn$^9$,  B. K. Gibson$^{10}$, E.K.\ Grebel$^{11}$, U. Munari$^{12},$ J. Navarro$^{13}$,  Q. Parker$^{14, 15}$,
\newauthor W. Reid, $^{16,17}$, G. Seabroke$^{18}$,  T. Zwitter$^3$ \\
$^{1}$Institute of Astronomy, University of Cambridge, Madingley Road, Cambridge CB3 0HA, UK\\
$^{2}$N\'ucleo de Astronom\'ia, Facultad de Ingenier\'ia, Universidad Diego Portales,  Av. Ej\'ercito 441, Santiago, Chile\\
$^{3}$Faculty of Mathematics and Physics, University of Ljubljana, Jadranska 19, 1000, Ljubljana, Slovenia\\
$^{4}$Department of Astronomy, Columbia University, 550 W 120th St, New York, NY 10027, USA\\
$^{5}$Leibniz-Institut für Astrophysik Potsdam (AIP), An der Sternwarte 16, D-14482 Potsdam, Germany\\
$^{6}$Laboratoire Lagrange, Université C\^ote d'Azur, Observatoire de la C\^ote d'Azur, Bd de l'Observatoire, CS 34229, F-06304 Nice cedex 4, France\\
$^{7}$Lund Observatory, Lund University, Department of Astronomy and Theoretical Physics, Box 43, SE-22100, Lund, Sweden\\
$^{8}$Observatoire astronomique de Strasbourg, Universit\'e de Strasbourg, CNRS, UMR 7550, 11 rue de l'Universit\'e, F-67000 Strasbourg, France\\
$^9$Sydney Institute for Astronomy, School of Physics, A28, University of Sydney, NSW 2006, Australia\\
$^{10}$E.A. Milne Centre for Astrophysics, University of Hull, Hull, HU6 7RX, UK \\
$^{11}$Astronomisches Rechen-Institut, Zentrum f\"ur Astronomie der Universit\"at Heidelberg, M\"onchhofstr.\ 12--14, 69120 Heidelberg, Germany\\
$^{12}$Dipartimento di Fisica e Astronomia Galileo Galilei, Universita' di Padova, Vicolo dell' Osservatorio 3, I-35122 Padova, Italy\\
$^{13}$ Department of Physics and Astronomy, University of Victoria, Victoria, BC, V8P 5C2 Canada\\
$^{14}$Department of Physics, CYM Building, The University of Hong Kong, Hong Kong, China\\
$^{15}$The Laboratory for Space Research, The University of Hong Kong, Hong Kong, China\\
$^{16}$Department of Physics, Macquarie University, Sydney, NSW 2109 Australia\\
$^{17}$University of Western Sydney, Locked Bag 1797, Penrith South DC, NSW 1797, Australia\\
$^{18}$Mullard Space Science Laboratory, University College London, Holmbury St Mary, Dorking, RH5 6NT, UK\\
}

\date{Accepted XXX. Received YYY; in original form ZZZ}

\pubyear{2017}

\begin{document}
\label{firstpage}
\pagerange{\pageref{firstpage}--\pageref{lastpage}}
\maketitle

\begin{abstract}
We apply the {\it twin method} to determine parallaxes to 232,545   stars of the RAVE survey using the parallaxes of Gaia DR1 as  a reference. To search for twins in this large dataset, we apply the t-stochastic neighbour embedding (\tsne) projection which distributes the data according to their spectral morphology on a two dimensional map. From this map we choose the twin candidates for which we calculate a $\chi^2$ to  select the best sets of twins.  Our results show a competitive performance when compared to other model-dependent methods relying on stellar parameters and isochrones. The power of the method is shown  by finding that the accuracy of our results is not significantly affected if the stars are normal or peculiar since the method is model free. We find twins for {60\%} of the RAVE sample which is not contained in TGAS or that have TGAS uncertainties which are larger than 20\%. We could determine parallaxes with typical errors of 28\%.  We provide a complementary dataset for the RAVE stars not covered by TGAS, or that have TGAS uncertainties which are larger than 20\%, with model-free parallaxes scaled to the Gaia measurements. 
\end{abstract}




\section{Introduction}

Stellar surveys are revolutionising our way of studying the structure and evolution of the Milky Way as large samples of stars allow us to investigate the properties of  their parent populations  in a statistical way. This revolution is now going through its climax thanks to Gaia and its first data release in September 2016 \citep{2016arXiv160904172G}.  Among this data release, the position, proper motions and parallaxes of more than two million of stars in the Tycho-2 catalogue became available (the so-called Tycho-Gaia Astrometric Solution (TGAS) sample). Of this sample, a good overlap can be found with spectroscopic surveys, such as LAMOST \citep{2012RAA....12..723Z}, RAVE \citep{2006AJ....132.1645S}, APOGEE \citep{2010SPIE.7735E..1CW}, GALAH \citep{2015MNRAS.449.2604D} and Gaia-ESO \citep{2012Msngr.147...25G}. 

The next data release will contain parallaxes of almost every target which has been observed with a spectrograph from the ground, allowing us to perform full chemodynamical studies of the Milky Way with unprecedented accuracy. At the same time, such complete datasets will help us to better calibrate the distance ladder by determining distances to field stars with different methods as  has been done with Hipparcos data \citep[][to name recent examples using spectroscopic survey data]{Binney14, 2014MNRAS.445.2758R,2015AJ....150....4C}.  

Most commonly, with the spectrum in hand, one can derive stellar parameters  which are used to place the star in the Herztsprung-Russell diagram and infer, with the help of stellar evolutionary models, the absolute magnitude of a star. Combined with apparent magnitudes, this is used to determine its distance modulus. The methods are usually tested using a sample of stars for which distances are known and applied to larger datasets for which distances are not known. In the above examples,  the methods have been so far tested against a set of Hipparcos stars, results from the literature using same approaches but developed by other groups as well as star clusters.   

The results for the distances obtained using stellar parameters and evolutionary tracks (a.k.a. spectrophotometric distances)  are, however, very model dependent \citep[see also discussion in][]{2013MNRAS.429.3645S}. On the one hand, accurate stellar parameters are required. Accuracy is obtained when we have a good handle of the systematic uncertainties in the modelling of stellar atmospheres and atomic data. We further need to have realistic  assumptions  to solve the radiative transfer equations and good control on other technical issues such as procedures to fit spectra  \citep[e.g.][]{2012A&A...547A.108L, 2016ApJS..226....4H, 2016arXiv161205013J}.  On the other hand, several parameters are needed to model the evolution of a star but cannot always be observationally constrained. Some examples include the mixing length, the overshooting parameters,  abundances of helium or the colours considered to fit the observables \citep[][or Miglio et al {\it in prep}, for a detailed comparison of different stellar tracks]{2015AJ....149...94H}.  It is a great challenge to use stellar models for distance determination and demonstrates the importance of Gaia data for improving our understanding of the many important physical processes governing stars. 

It might be possible to overcome the aforementioned limitations in our  modelling of the  structure and evolution of  stars and still use stellar survey spectra to determine distances accurately.   \citet[][hereafter J15]{Jofre15}  recently introduced {\it the twin method}, which relies on the assumption that if two stars have identical spectra { within the observed wavelength range},  it is possible to determine the distance of one star provided the distance of the other star is known and both stars  have photometry with the same passbands. This is possible  because of the proposition that if the stars have the same spectra it implies that they must also have the same intrinsic luminosity and therefore, the difference of their brightness in the sky is directly related to their difference in distance. The method is model free, as the spectra are  used as a way to assess that both stars have the same intrinsic luminosity, but nothing else.  This idea of spectroscopic twins can indeed be applied not only to determine distances to stars, but also to galaxies \citep{1984ApJ...282..382P} or to supernovae \citep{2015ApJ...815...58F}. 

J15 demonstrated the first application of the twin method on a set of Hipparcos stars and  open clusters  using high-resolution and high signal-to-noise ratio (SNR) spectra from the HARPS instrument located in La Silla, Chile \citep{2003Msngr.114...20M}. It was shown that with twin stars it is possible to determine parallaxes to 10\% precision for  FGK-type stars. This precision is  the order of magnitude of the Hipparcos parallax uncertainties used for that study. The excellent performance of the method on that kind of data was then further demonstrated in the determination of the distance of the Pleiades (also called Melotte 22) by \citet[hereafter M16]{2016A&A...595A..59M}. That measurement has since been corroborated by Gaia DR1 \citep{2016arXiv160904172G}. \\

Here we explore the twin method further by applying it for the first time to survey spectra. We choose to use the RAVE survey due to its large overlap with TGAS. Furthermore,  its spectra cover roughly the same wavelength range as Gaia RVS spectra.  We want to address the following questions  in this analysis: 
\begin{enumerate}
\item {\it How accurate are our results  when applied to a spectroscopic survey of considerably lower resolution and wavelength coverage than HARPS? }
\item {\it For how many stars in a spectroscopic survey can we apply the twin method to determine distances and how to the results depend on spectral type? }
\end{enumerate}
Question (i) is important in the era of Gaia since climbing the cosmic ladder with twin stars will have the largest impact for very faint stars, where a high resolution and high SNR spectrum is difficult to obtain and direct measurement of parallaxes becomes uncertain. Question (ii) addresses  the usefulness of the method. While the method is model-free, it requires having a twin star with well-measured parallax and  observed with the same photometry and spectrograph.  Also, it helps us to know how good can be the results for stars observed in different evolutionary stages, especially for serendipitous peculiar stars. Finally, this analysis serves as test case for the future use of twins in Gaia spectra. 

As a result of this analysis we do not only provide answers for the above questions, but also a catalogue of parallaxes for the RAVE stars that are not in TGAS. Such parallaxes should help with chemodynamical analyses of the Milky Way. It is worth to mention that new UCAC5 proper motions \citep{2017AJ....153..166Z} are also available for all the RAVE stars, which have uncertainties comparable  than TGAS proper motions.  Therefore, once combined with the twin parallaxes, we can have the motion of the stars through space in 6 dimensions with TGAS quality for a much large dataset within RAVE  before Gaia DR2 becomes public. In \sect{data} we briefly describe the RAVE and TGAS data used in this work and in \sect{method} we explain the method employed to find twins and determine parallaxes. In \sect{results} we discuss the performance of our results with respect to Gaia parallaxes as well as results obtained by other methods. In \sect{catalogue} we describe the catalogue of twin parallaxes for the non-TGAS stars and in \sect{conclusions} we summarise and give our conclusions.

\section{Data}\label{data}
In this work we use the fifth data release of the RAVE survey (RAVE DR5) which is described in detail in \citet[][hereafter K17]{2017AJ....153...75K}. Briefly, the dataset contains  about 500,000 spectra of resolution  $R\sim7500$ and wavelength coverage  8410-8795\AA\ observed between  2003 and 2013 for which accurate radial velocities (RVs) are available. In this sample,  about 200,000 stars have parallaxes and proper motions from the TGAS sample in Gaia DR1 \citep{2016arXiv160904172G}. 

The RAVE spectra do contain information on the temperature, gravity and metallicity of the observed star. We have measurements of stellar parameters at our disposal. The RAVE-collaboration DR5 parameters which were determined using the RAVE DR4 pipeline \citep{2013AJ....146..134K}  with improvements thanks to  calibrations using Kepler 2 targets for seismic gravities \citep{2016arXiv160903826V}, the Gaia benchmark stars spectral library \citep{2014A&A...566A..98B} and stellar parameters \citep{2015A&A...582A..49H, 2014A&A...564A.133J}, as well as other results obtained from high-resolution studies and { temperatures obtained with infrared flux method}. From the stellar parameters, model-dependent distances are available which we can use to compare our results. While current distances are those obtained following the method described in \cite[][hereafter B14]{Binney14}, there are several previous works that have provided with important lessons needed for this task \citep{2010A&A...511A..90B, 2010A&A...522A..54Z, 2011A&A...532A.113B}. { For discussion on how these improved parameters affect the spectroscopic distances from RAVE we refer the reader to the extensive discusison in Section 9 of K17. Here we note that we use the parameters only for assessing the performance of our method for different kind of stars, but the parameters are not an input of our method. }

Recently, a second set of RAVE stellar parameters became available,   provided by \cite{2016arXiv160902914C}, who employed {\it The Cannon} data-driven approach \citep{2015ApJ...808...16N}, based on training the data using parameters from results obtained from data with higher resolution and wavelength coverage than RAVE. This set of parameters is referred to RAVEon.  We emphasise that the information on the stellar parameters is not used in our case to determine distances, but only used to assess the accuracy and reliability of the twin method. 

Since the distances are determined by looking at the difference in the observed magnitudes of the stars, we use the photometric data coming from 2MASS \citep{2006AJ....131.1163S} which are available for all  the stars in the sample.  In addition to the photometry,  we need to have a set of reference stars for which we already know the distance.  These come from the stars that are contained in TGAS. 


\section{Method}\label{method}
Our method is based on selecting stars that are similar based on the morphology of their spectra.  This follows the principle used in J15 and M16 but here we employ a new approach for two main reasons. First, the RAVE spectra are of considerably lower resolution and wavelength coverage than the HARPS spectra used in the previous works (R = 115,000 and a wavelength coverage from 400 to 600~nm approximately). This made it impossible to concentrate on individual absorption lines, especially when these lines were those selected to be clean lines containing information on the metal content and structure of a wide variety of FGK-type stars according to \cite{2015A&A...582A..81J}. The vast majority of these lines are not part of the RAVE spectral domain and therefore cannot be used. An approach aimed to consider the entire shape of the spectra is more suitable for this case.  

 A second reason is the size of the data set. In J15 the method was applied to a set of $\sim$700 Hipparcos stars while in  M16  the distance of the Pleiades was determined by looking for twins of $\sim$ 20 cluster members only using the reference set of $\sim$700 Hipparcos stars. Here we aim to analyse the entire RAVE survey, that is, approximately 500,000 stars, which is several orders of magnitudes larger. Finally, our reference set of stars for which parallaxes are known is larger now that the first data release of Gaia is available. In our case, the TGAS  set of stars is of about 250,000 stars. A technique that can handle this large dataset is needed.

\begin{figure*}
	\centering
	\includegraphics[width = \textwidth]{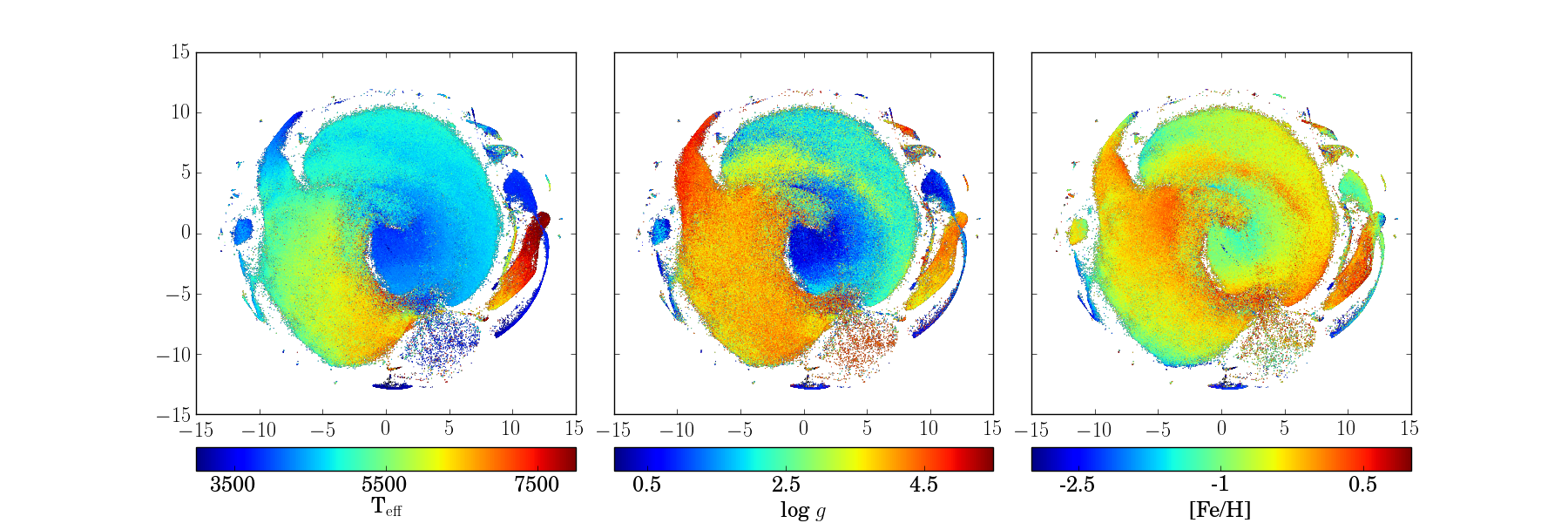}
    \caption{\tsne\ maps of the RAVE stars coloured by effective temperature (left), surface gravity (middle) and metallicity (right). The parameters are those of the RAVE DR5 pipeline. Black colour corresponds to stars with no parameters in DR5, which are mostly those with very low SNR.}
    \label{rave_tsne}
\end{figure*}


In this era of big data, such techniques exist on the market, and here we chose the machine-learning method called ``t-Student Stochastic Neighbour Embedding"  \citep[\tsne, ][]{maaten2008visualizing}. The method converts a high-dimensional dataset into a matrix of pair-wise similarities which starts by converting the high-dimensional Euclidean distance between datapoints into probabilities that represent similarities. A final two-dimensional (2D) map is produced which aims to visualise the  data into the dimensions in which the data are distributed by their  similarities by clustering similar features together. 

 \tsne\ has already been applied to surveys of stellar spectra with the purpose of identifying potential peculiar stars or  technical issues such as data reduction problems.  For details on this we refer the reader to \citet[][hereafter T17]{2017ApJS..228...24T}, who recently presented the method applied to GALAH spectra, showing on the one hand how the stars separate well in this map according to their stellar parameters, and on the other hand how peculiar objects lie in definite regions in the map.   We also note that the \tsne\ 2-D map applied to the RAVE spectra was used by \cite{2016arXiv160903826V} to highlight that one part of the map corresponds to the giants while the other part to the dwarfs and recently in \cite{2017arXiv170405695M} to detect metal-poor stars. 

 The motivation for applying \tsne\ to spectra comes from the already successful attempt to reduce dimensions and study the morphology of spectra using the neighbour embedding method presented in \citet[][hereafter M12]{2012ApJS..200...14M}. They showed that the normal  and the peculiar spectra are located in different parts in a 2-D map. 
  That method has become the inherently quantifiable way of assigning flags to peculiar RAVE spectra, which are basically assigned by looking at the closest neighbours of given templates with known peculiarities. Recently, an extensive analysis of RAVE stars with chromospheric activity detected with this method has been published \citep{2016arXiv161202433Z}. 
  
The method presented in M12 however suffers from a crowding problem, meaning that it squeezes the ``normal" spectra to the central part of the map, so in the particular case of analysing twin stars,  it is not as good as \tsne.  Thus, for the purpose of finding twin stars, we employ the same procedure as in T17 but adjusting the parameters to cluster similar ``normal" stars as much as possible. That is, we computed the \tsne\ embedding by  (1) projecting our dataset into two-dimensional space, (2) setting the Barnes-Hut parameter $\theta$ to 0.5, and (3) setting the perplexity parameter to 10 which produced in general sparser projection maps with denser collections of points than in T17. More  details of how this is done in RAVE data can be found in \cite{2017arXiv170405695M}.

In \fig{rave_tsne} we show the \tsne\ maps coloured by the DR5 stellar parameters. It is possible to see how the different kinds of stars are distributed in the map with stars with the same parameters close together.  This figure also validates the quality of the DR5 stellar parameters.  As in T17 and \cite{2017arXiv170405695M}, a large part of the data are ``normal" and populate the largest regions in the map. There are however some islands, produced notably by  very cool giants or by hot stars. 

\begin{figure}
	\centering
	\includegraphics[width = \columnwidth]{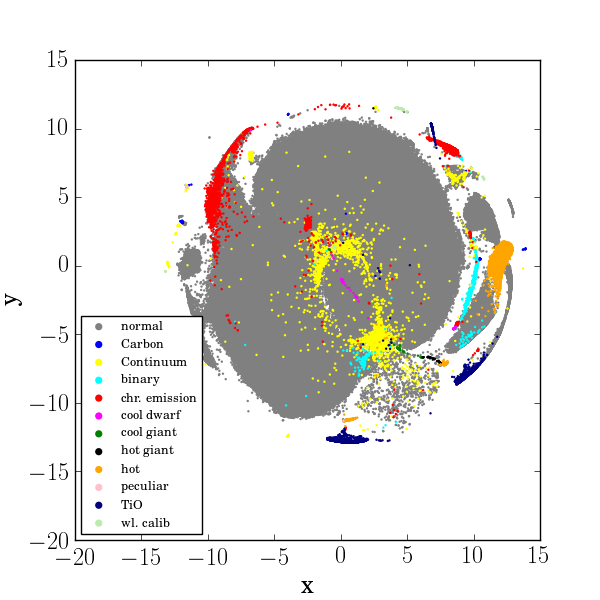}
    \caption{\tsne\ map of the stars which have $c1$, $c2$ and $c3$ flags consistent with the nomenclature of the legend. The majority of the stars are coloured with grey, indicating they have normal spectra. The analysis of the flags of these data is described in M12. }
    \label{flags_tsne}
\end{figure}

The same \tsne\ map is  shown in \fig{flags_tsne} coloured by the different flags employed for the $c1, c2, c3$ variables in DR5, which correspond to the morphological classification of M12. Here we plot the stars whose 3 first morphological classification flags are identical. We can see how the central region of the map is mostly grey (normal stars) but there are some stars that have been flagged.   This is consistent with the study of the morphology of RAVE spectra of M12 who concluded that about 90\%-95\% of the stars are normal, and the rest are either peculiar (double-line binaries, chromospherically active, strong TiO bands, etc) or have some technical failure in the spectra (ghosts, wavelength calibration, etc).  Each of these not-normal stars have a flag denoted by different colour. 

In the \tsne\ map,  the differently flagged stars populate well-defined regions and are consistent with our expectations from the stellar parameters. For example,  some islands of cool giants  are flagged with strong TiO bands (dark blue colour). There is also a flag for the very hot main-sequence stars (orange colour). Binaries populate an island coloured in cyan and chromospherically active stars populate the edge of the main island, as well as the binary island, and are coloured with red. Stars with problems  in continuum normalisation are coloured with yellow and are distributed in different parts of the map but still clustered together in three main groups, mostly representing the cool stars. This is not surprising, we know that cool stars have strong absorption features making it more difficult to find the continuum in the spectra \citep[see e.g.][for extensive discussions]{2012A&A...547A.108L, 2014A&A...564A.133J}. It is worth to mention that this normalisation issue does not necessarily impact the DR5 parameters as the pipeline re-normalises the spectra to account for such problems.

Both figures tell us that the \tsne\ map indeed orders the stars according to their morphology and thus, neighbouring stars in this map can very likely be treated as twins. In order to use this map and the twin method to determine parallax to the non TGAS-RAVE stars we basically search for stars in the TGAS-RAVE set  that are neighbours of the non TGAS-RAVE stars. To do so, we first review the basics of how we determine the parallax of twin stars below.  

\subsection{Determination of twin parallaxes}
First, we emphasise that in this work we determine parallaxes and not distances. This is because we use the parallax of TGAS as a reference. To transform parallax to a distance, we refer to the discussion of \cite{2015PASP..127..994B}  who highlights  the challenges of inverting and estimating the parallax, in particular its uncertainty. 
We follow the procedure of J15 and M16 to determine the twin parallaxes.  Here we consider the 2MASS $J$ and $K_s$ photometric bands for two main reasons. First,  they are less affected by extinction and second, all stars in the RAVE catalogue have this photometry.  

An important assumption to mention here is that spectroscopic twins must not only have the same luminosity, but also the same intrinsic colour. Therefore the extinction between two stars can be directly related to the differences in their observed colours, $(J-Ks)_1$ and $(J-Ks)_2$.  This allows us to use the following equation to determine the parallax:
\begin{equation}\label{eq1}
K_{s,1} - K_{s,2} - \frac{R_K}{R_J-R_K} [(J-K_s)_1 - (J-K_s)_2]  = -5\log(\varpi_1/\varpi_2),
\end{equation}

\noindent  { where $R_J=0.7$ and $R_K=0.3$  are the ratio of total-to-selective extinction in the $J$ and $K_s$ bands, respectively.} The values are taken from the empirical transformations done by \cite{2013MNRAS.430.2188Y}.  The importance of assuming that twin stars have the same colour is that we do not need to correct the photometry of the stars  by extinction, which implies that we do not depend on using dust maps such as those provided by \cite{1998ApJ...500..525S} or \cite{2011ApJ...737..103S}. We note that Eq.~\ref{eq1} is only valid if the assumption that $d = 1/\varpi$ is good. This is only the case for precise parallaxes  \citep{2015PASP..127..994B}.

\begin{figure}
	\centering	
	\includegraphics[width = \columnwidth]{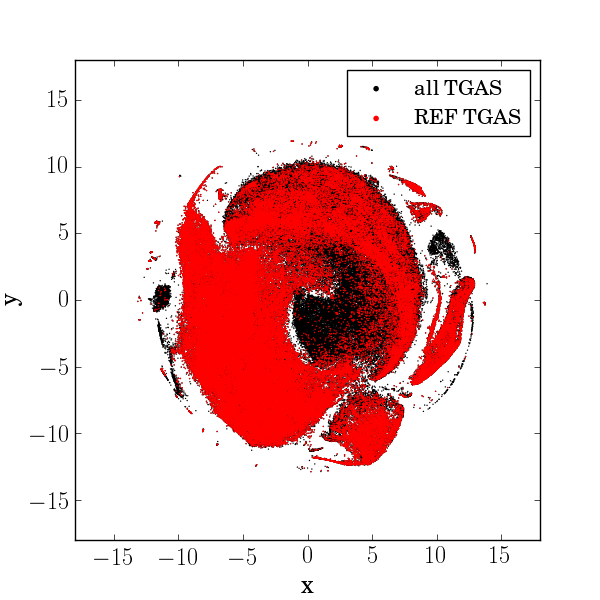}
    \caption{\tsne\ maps for the TGAS sample and the sample with accurate and positive parallaxes, which is the one suitable for determination of twin parallaxes. A large number of stars is rejected from this cut, in particular cool distant giants.}
    \label{tgas_good}
\end{figure}

In \fig{tgas_good} we show the distribution of stars in the \tsne\ map of the entire TGAS sample in black, and for accurate (better than 20\% fractional uncertainty) and positive parallaxes from TGAS with red.  About half of the RAVE-TGAS sample is suitable as reference, covering large parts of the parameter space except the coolest giants, which include the most distant stars because they are very bright. The performance of our method for such stars will have to be investigated when the second data release of Gaia becomes public in 2018.

To get a robust result in the twin parallax, we require to have at least  5 twin candidates with a measured parallax of accuracy better than 20\% of a given target (e.g. ``quintuples'') to compute a distribution of twin parallaxes following Eq.~\ref{eq1} \citep[see below and  in M16, as well as discussions in ][]{astroMLText}. We define the final parallax to be mean of this distribution after performing a $5 \sigma$ clipping to remove spurious values due to wrong reference parallaxes, poor photometry or bad spectra, and its uncertainty is the standard error of this mean.   We note however that this value corresponds to the internal error only.  To have a realistic measure of the uncertainty of our parallax we need to account for external errors. For that we study the agreement in the parallaxes derived with external sources which is discussed below. 
We finally comment that this work is done only with reference stars with positive parallaxes, as we can only determine $\varpi_2$ by applying Eq.~\ref{eq1} if $\varpi_1$ is positive.

 \begin{figure}
	\centering	
	\includegraphics[width = \columnwidth]{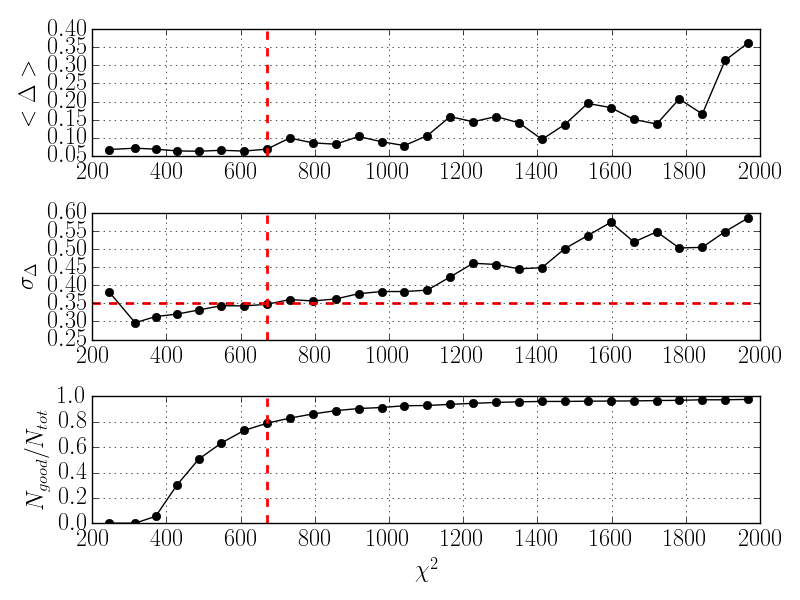}
    \caption{Distribution of $\chi^2$ of pairs of a randomly selected sample of 1,000 stars. Upper panel: the mean difference of twin and TGAS parallaxes. Middle panel: the dispersion of the difference between twin and TGAS parallaxes. Lower panel: The fraction of stars in the sample for which we could determine twin parallaxes following the criteria discussed in the text. The dashed horizontal red line indicates the external accuracy of 35\%. }
    \label{chi2}
\end{figure}

\subsection{Twin stars suitable for parallax determination}\label{twinicity}

To take the best twins of a given target from the \tsne\ map we choose the stars within a circle of radius $R$ around the location of that target in the map. That gives us the stars with closest spectra and therefore the best twin candidates for each star. We then selected the best candidates using a $\chi^2$ criterion which would minimise the fractional difference between the twin parallax and the direct value by TGAS which we define by.
\begin{equation}\label{delta_frac}
\Delta = \frac{\varpi_{\mathrm{TWIN}}-\varpi_{\mathrm{Gaia}}}{\varpi_{\mathrm{Gaia}}} \,
\end{equation}
\noindent where $\varpi_{\mathrm{TWIN}}$ and $\varpi_{\mathrm{Gaia}}$ correspond to the twin and Gaia parallax of an object respectively.

The aim is to have a radius $R$ which encloses at least 5 twin candidates with accurate parallaxes and provides the distribution of $\Delta$ with a mean that is closest to zero and small dispersion. The dispersion is attributed to the fractional external uncertainty of our method.  Furthermore, we require a value of $R$ which maximises the number of twin candidates, i.e. we aim to determine parallaxes for the entire dataset.  In order to find this value, we  randomly selected a sample of 75\% of the TGAS-RAVE stars as reference stars. The parallaxes of these stars were used to measure the twin parallax of the remaining 25\% of the stars (30,303 stars), which were then compared to the Gaia parallaxes.   We note that these numbers correspond to stars which were selected to have TGAS parallaxes that were measured with an uncertainty of better than 20\% and were positive. 

In this sample,  30,143  had at least 5 twin candidates enclosed in a circle with a radius of $R=0.5$ with the median number of twin candidates being  341. For these candidates we calculated their $\chi^2$ with the target star and studied the distribution of all pairs in order to find which $\chi^2$ would give us the best agreement with TGAS for the largest number of RAVE targets.  

In order to calculate the $\chi^2$, we had to normalise the spectra, correct them to have in the same reference frame, and finally re-sample them such that every pixel could be compared. This was done using the functionalities of {\tt iSpec} \citep{2014A&A...569A.111B}. Since we do not aim to compare the spectra with models, we performed a cross-correlation of the twin candidates with the desired target directly. The result of this cross-correlation was then used to align and re-sample the twin candidates to the reference frame of the target. Likewise, the spectra of the twin candidates were normalised using the target as the reference template.  { While this procedure was more time-consuming that correcting by radial velocity and fitting  with polynomial fits the continuum of the spectra, this differential comparison gives us the chance to resample one spectrum to the reference template which gives us an {accurate} goodness-of-fit measure. We further comment that with this procedure we could use the the peak and width of the cross-correlation to select the best candidates, which were expected to have the highest peak and narrower width.  Since all stars selected to belong to the circular area of the \tsne\ map and were thus quite similar to the template already, the two cross-correlation output values were not clearly  indicative as the $\chi^2$ as we see below. }  

\begin{figure*}
	\centering
	\includegraphics[width = \textwidth]{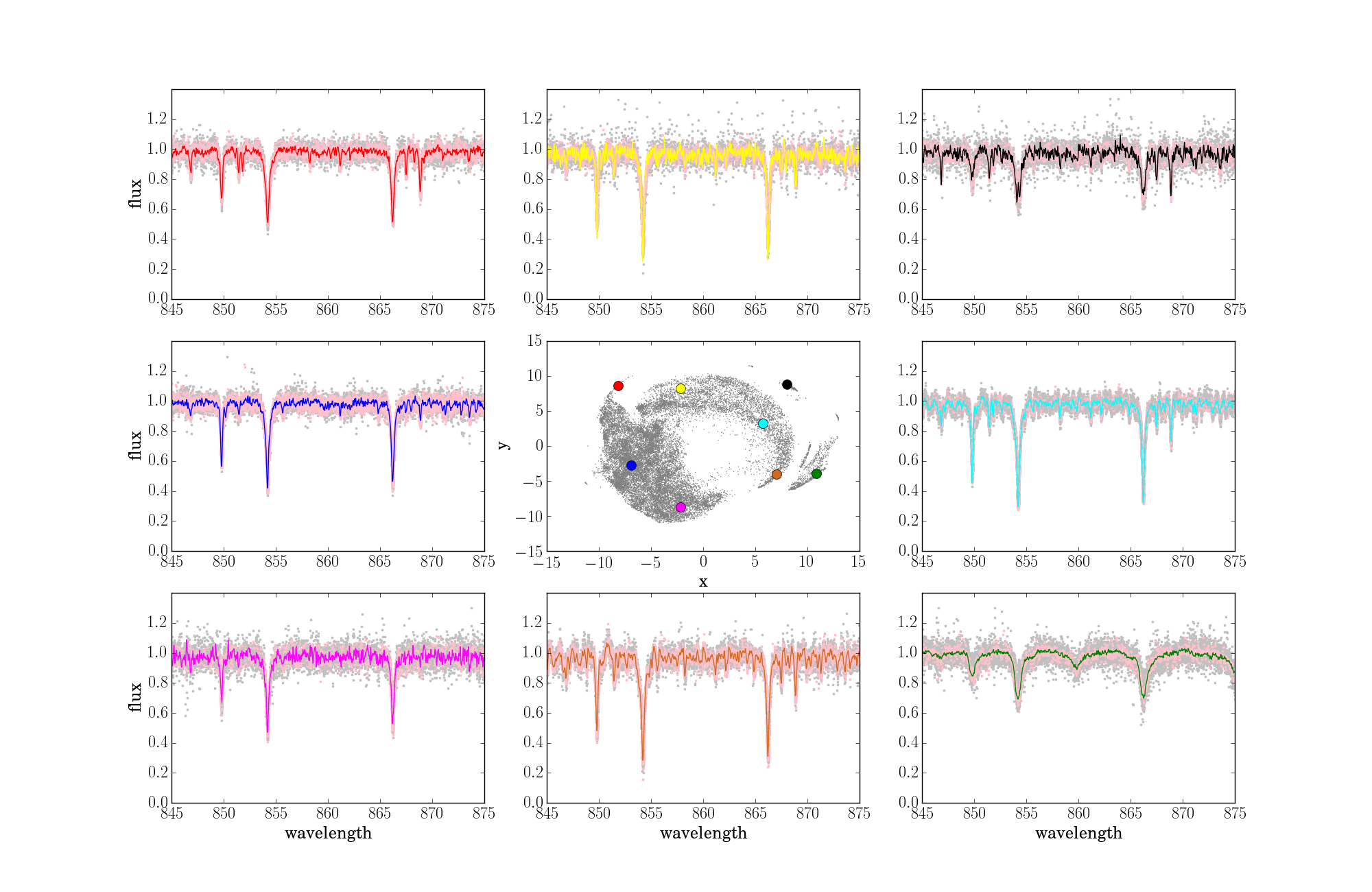}
    \caption{Example of RAVE spectra of different types. Their twin candidates with $\chi^2>670$ are plotted with grey while the twin candidates with $\chi^2<670$ are plotted with pink. It is possible to idenfity the position of the stars in the \tsne\ map by identifying the circle of the same colour as the spectra in the central panel.  }
   \label{twin_spectra}
\end{figure*}

To find an appropriate cut in $\chi^2$, we randomly selected 1,000 stars and studied their $\chi^2$ distributions as a function of the agreement between the derived twin parallax and TGAS. This can be seen in \fig{chi2}, which shows how different quantities depend on $\chi^2$. For this plot,  we binned the sample in $\chi^2$ and calculated the mean and the width of the $\Delta$ distribution. The width was defined as the difference of the 25\% and 75\% percentiles \citep{astroMLText}. These quantities  are shown for each $\chi^2$ bin on the top and middle panel, respectively.   On the bottom panel we show the fraction of the stars in our sample for which a twin parallax could be determined after requiring at least 5 twin candidates that have $\chi^2$ of the corresponding bin. 

There is a correlation of $\chi^2$ as a function of agreement between TGAS and our results, in which the larger the $\chi^2$, the larger the disagreements.  However, by forcing $\chi^2$ values that are too small, we compromise too by obtaining results for fewer stars. By requiring to have at least 5 twin candidates which have $R < 0.5$ in the \tsne\ map, have parallaxes more accurate than 20\% and individual twin parallaxes in 35\% agreement with TGAS, we need to consider twin pairs which can not have a  $\chi^2$ above 670. This allows us to determine twin parallaxes for 80\% of the TGAS sample. 

{To illustrate this we show in \fig{twin_spectra} examples of spectra of stars that are located at different parts of the \tsne\ map, which is plotted in the central panel for reference. Each spectrum is represented with a different colour and we see how their morphology is different at different regions in the map.  In addition, the spectra of all twin candidates which have $R<0.5$ from the target have been added with a scatter plot to the corresponding panels. Stars with $\chi^2 > 670$ are represented with grey while stars with $\chi^2 < 670$ are illustrated with pink. It is possible to see that the pink stars are systematically more alike to the target than the grey stars.  }

The final twin parallaxes (i.e. the average of all twins of a given target which satisfy the conditions of above) have an agreement with TGAS of 28\% and a bias of 9\% towards larger values in the case of the twin parallaxes. The fact that the agreement is better than 35\% is because we have included all stars which have smaller $\chi^2$ than 670, which show to have better agreement than 35\%.

 \begin{figure}
	\centering	
	\includegraphics[width = \columnwidth]{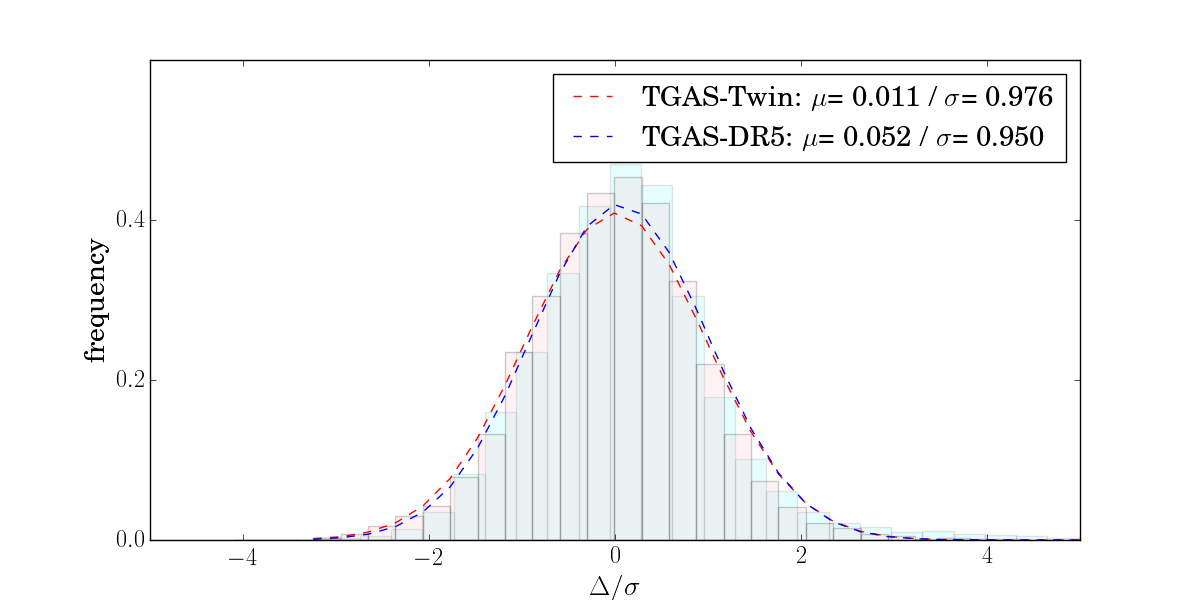}
    \caption{Upper panel:  Distribution of $\Delta/\sigma$ between our results and TGAS. Lower panel: distribution of fractional uncertainties (internal and external) for the Twin parallaxes and for TGAS for the same stars.}
    \label{distr_uncertainties}
\end{figure}

\subsection{Uncertainties}\label{uncertainties}
To see if our uncertainty estimates are well-determined we consider the difference of the twin and Gaia parallax of a given star as follows: 
\begin{equation}\label{delta_def}
\Delta / \sigma= \frac{\varpi_{\mathrm{TWIN}}-\varpi_{\mathrm{Gaia}}}{\left (\sigma(\varpi_{\mathrm{TWIN}})^2+\sigma(\varpi_{\mathrm{Gaia}})^2 \right)^{1/2}} \ ,
\end{equation}
\noindent where $\varpi_{\mathrm{TWIN}}$ and $\varpi_{\mathrm{Gaia}}$ correspond to the twin and Gaia parallax of an object respectively, and  $\sigma(\varpi_{\mathrm{TWIN}})$  and $\sigma(\varpi_{\mathrm{Gaia}})$ are their respective uncertainties.   Here $\sigma(\varpi_{\mathrm{TWIN}})$ corresponds to the sum of the quadratures of the internal uncertainty (standard error of mean parallax of all twins) and external uncertainty (agreement with TGAS of 28\%, see above). Expression~\ref{delta_def} is taken from B14 and serves to indicate if the differences are significant considering the uncertainties of the measurements.

In \fig{distr_uncertainties} we show the distribution of $\Delta/\sigma$. We obtain a small offset of $\langle  \Delta/\sigma \rangle = 0.01$ and unit dispersion, indicating that our uncertainties are well determined.  
 For comparison we also show with blue the $\Delta/\sigma$ distribution of the parallaxes reported as part of RAVE-DR5 which are determined as described in B14. We note that the distances  used here are based on RAVE DR5 stellar parameters. The distribution of DR5 parallaxes has a similar offset and dispersion as our distribution.  


We comment that systematic uncertainties in the TGAS sample are extremely complex as they are highly correlated  with the position of the star in the sky (the scanning law), the distance of the stars, spectral type, and so on \citep{2016A&A...595A...4L, vL17}.  Extensive discussions have recently arisen in the literature concerning systematic uncertainties of TGAS , which not necessarily agree \citep[e.g.][to name a few]{2016ApJ...832L..18J, 2016arXiv160906315G, 2017A&A...598L...4D, vL17, 2017A&A...599A..67C, 2017arXiv170508988H}. Therefore, in this work we only take the uncertainties reported directly by the first data release of Gaia, which are described in \cite{2016A&A...595A...4L}. Further systematic uncertainties will be thus propagated through the entire sample according to Eq.~\ref{eq1} (see  also J15 for discussions).


\section{Results}\label{results}

In this section we investigate how our results for the TGAS-RAVE sample agrees with the Gaia values, as well as well as with the model-dependent results provided as part of RAVE-DR5 which are calculated using the procedure described in B14. We also compare our results with open clusters.

\begin{figure*}
	\centering
	\includegraphics[width = \textwidth]{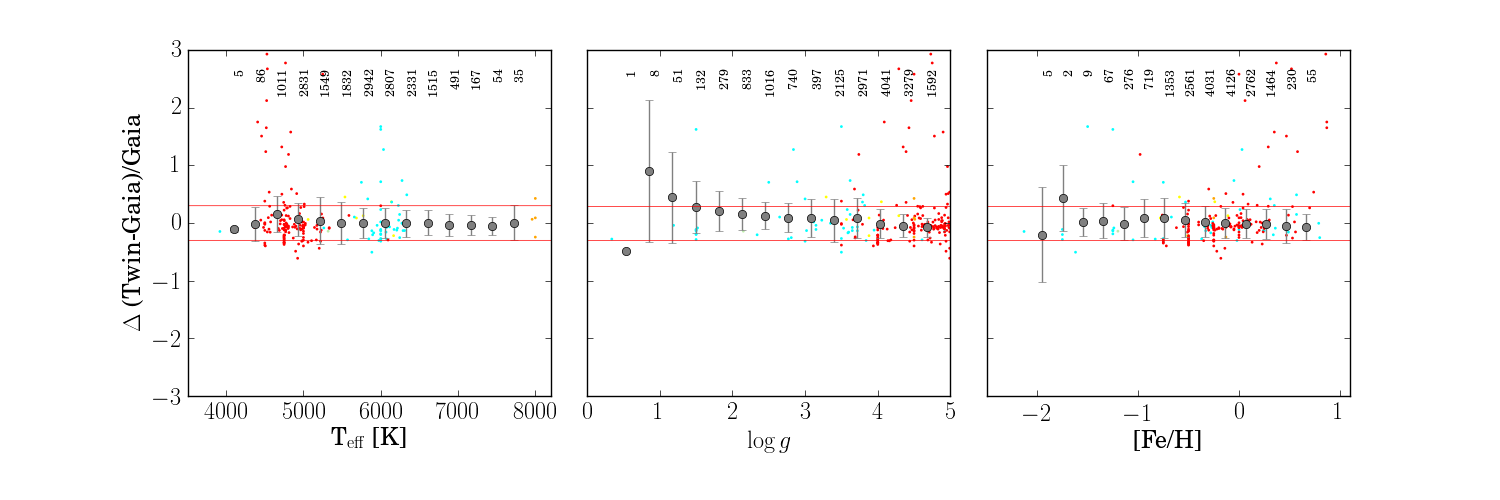}
	\includegraphics[width = \textwidth]{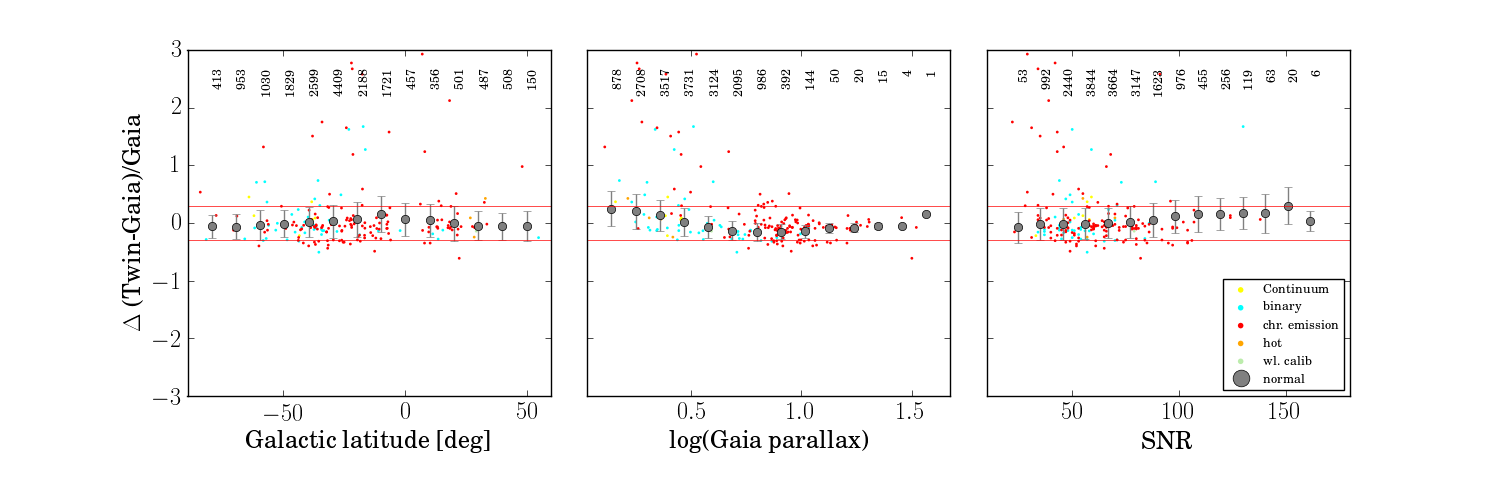}
   \caption{Upper panels: $\Delta$ comparisons of parallaxes between TGAS and our results as a function of stellar parameters taken from RAVE DR5.    Lower panels:   comparison of parallaxes as a function Galactic latitude, parallax, and SNR.  For better visualisation, normal stars have been binned with their mean of $\Delta$ represented with the cicle and the dispersion with the error bar.  The number of stars in each bin is indicated at the top of each panel. }
    \label{plx_params}
\end{figure*}

\subsection{Comparisons of parallaxes}\label{comparisons}

We consider only the randomly selected 25\% (see above) of the RAVE-TGAS  sample with uncertainties above 20\% and positive parallaxes to see with what accuracy  we can determine the distances of twin stars when RAVE spectra are available. From the 30,303 selected stars,  we could determine parallaxes for about { 80\%} this sample when adopting a value of  $R=0.5$  for the radius enclosing the twin stars of a given target and a $\chi^2 \leq 670$, which corresponds to a total of 22,927 stars.  
 We compare the Gaia parallaxes with our parallaxes as well as with parallaxes determined considering stellar parameters and isochrones. 
 
When considering (model-dependent) spectrophotometric distances, parallaxes could be determined for  { 22,028} stars with RAVE-DR5 parameters and for {21,052} stars with RAVE-on parameters.  In both cases, an additional criterion based on the quality of the parameters had to be adopted to provide a distance. In the case of the distances reported in B14,  no distances for stars whose parameters were too far from the isochrones employed could be provided. We found that stars with DR5 parameters having temperatures above 7500~K and metallicities below $-3$ had no DR5 parallax provided.  In the case of the distances derived from RAVE-on parameters, an initial selection was done to the stars removing those stars whose parameters were not reported and those with $rchi2>3$, which indicates a low quality of the fits.   Like in the case of B14, no further distances could be provided when the parameters were too far from isochrones. We found that most of the low SNR stars suffered from these selections, and that no star with $\mathrm{SNR}<10$ had a distance determined from RAVE-on parameters. 
  
   Low SNR stars have been flagged for problems with continuum level by M12, making our criteria to assess good twins more uncertain. Thus, for low SNR our twin parallaxes also became very uncertain.  To compare the performance of our results with the model-dependent results, we compare parallaxes obtained for the exact same sample, i.e, for the sample with derived twin parallaxes and spectrophotometric parallaxes (considering both, DR5 and RAVE-on parameters).  Our sample thus comprises of {21,052 stars. }

\subsubsection{Twin versus Gaia parallaxes}

 The fractional difference  of the Gaia parallaxes relative to our values, as defined in Eq.~\ref{delta_frac},  as a function of stellar parameters is displayed in the upper panels of \fig{plx_params}.  In this comparison we use the RAVE DR5 parameters for reference for the x-axis, but using the RAVE-on parameters do not change the main features of these figures as in general both parameters agree (see K17 for discussion). We recall here that our method is independent of the parameters of the stars, hence here they are displayed for analysis of how well we can obtain parallaxes for different kinds of stars only.  In the lower panels we show the differences of parallaxes as a function of Galactic latitude, as a function of parallax (in logarithm), as well as a function of Gaia parallaxes and  SNR of the spectra. In all panels we plot with grey the normal stars and with colours the flagged ones (see \fig{flags_tsne}). We binned the data and calculated the mean and width of the distribution for the normal data for each bin; the numbers of stars in each bin are indicated at the top of each panel. These are represented with the filled grey circles and the error bar.

\begin{figure*}
	\centering
	\includegraphics[width = \textwidth]{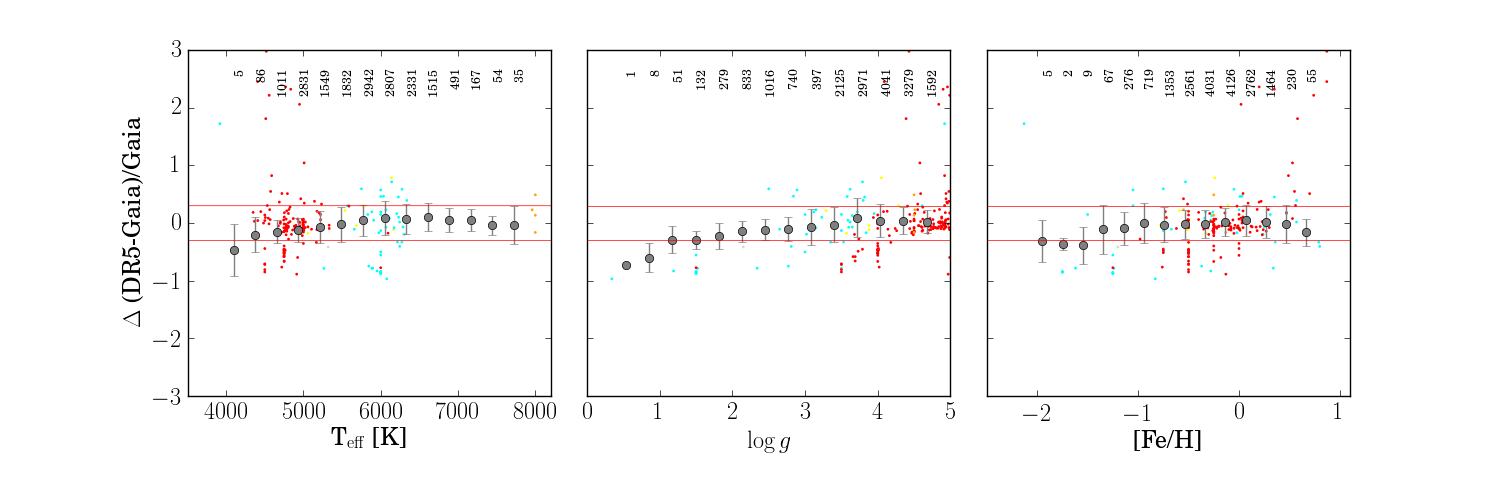}
	\includegraphics[width = \textwidth]{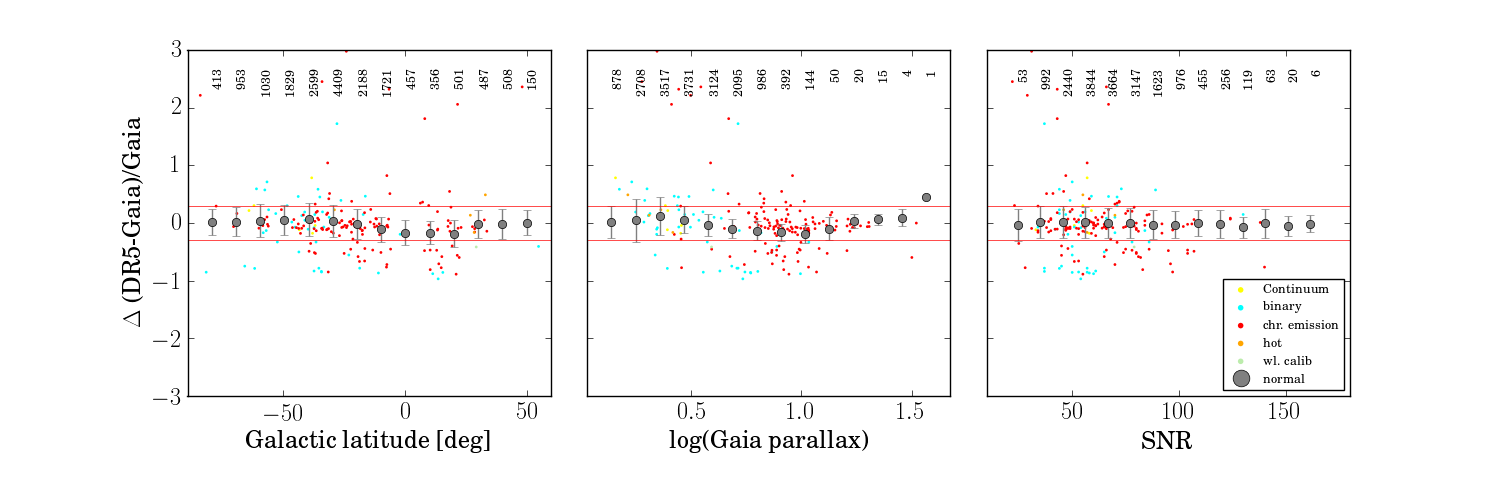}
   \caption{Upper panels: Same as \fig{plx_params} but here we compare our parallaxes with the model-based ones \citep[see][for details]{Binney14}. Further quality flags regarding the convergence of the DR5 pipeline have not been taken into consideration for this sample. Red horizontal lines represent 30\% difference. }
    \label{plx_params_dr5}
\end{figure*}

The scatter between Gaia and our parallaxes decreases with increasing temperature, in which stars with temperatures below 5000~K have a dispersion that can be larger than 30\%. These stars are located in a region of the \tsne\ map which overlaps with dwarfs.  By requiring twins to be those enclosed within a given radius $R$ in the map, we might be using wrong assumptions of the absolute magnitude, i.e., confusing our giants with dwarfs and hence, underestimating their distances. This is consistent with the trend seen for surface gravity. For giants we obtain an overestimation of parallaxes and a very large scatter, because using the magnitude of a dwarf as a reference for a giant implies closer distance for the giant.  This reflects the difficulty of separating between dwarfs and giants when only spectra around the Ca~II triplet are available \citep[e.g.][]{2011A&A...535A.106K, 2016A&A...585A..93R, 2016arXiv160903826V}.  We also note that by selecting only stars with accurate Gaia parallaxes, we have a sample more dwarfs than giants, which can be seen by the number of stars in each bin.  For stars with surface gravities above 2, the scatter of $\Delta$ is of the order of 30\% and there is no significant offset.  

For metallicity we do not obtain a clear trend of $\Delta$, except that for metal-poor stars our method becomes more uncertain. We note however that there are very few stars at low metallicities, which could be the cause of the apparent large error bars. The agreement as a function of Galactic latitude also does not show an evident trend, although at $b\sim-5^\circ$ a slight increase of parallaxes is found. This could be the reflection that employing the ratio of total to selective extinction ratio of $R_K=0.3$ is not the best approximation for very extincted regions.  This slight difference is however still negligible compared to other uncertain regions in the parameter space such as the low gravity stars. 

There is a dependency of $\Delta$ with Gaia parallax, in the sense that for smaller Gaia parallaxes a larger twin parallax is found. Again that might be related to the giants, which due to the selection effect of RAVE stars, have a tendency to be further away and thus have smaller parallaxes. These are uncertain with a systematic offset towards larger twin parallaxes, which is what we find here. For larger parallaxes (of the order of 10 mas or more) our parallaxes have a very good agreement with Gaia, with very small dispersion.  Finally, our results do not have a strong dependency on SNR. $\Delta$ slightly increases with SNR but remains within 30\% agreement with a scatter that seems to be the same in every bin.


In general, we obtain parallaxes with the same accuracy for normal and for flagged stars, demonstrating our ability to determine parallaxes independently of stellar parameters. It is encouraging to note that we retrieve accurate distances for most of the chromospherically active stars (red), carbon stars (light blue), hot stars (orange), TiO stars (blue), and binaries (cyan). In many cases there are numerous normal stars within the radius $R$ of the star flagged (see \sect{twinicity}), i.e. the signatures of peculiarity are in these occasions very weak.     This suggests that this peculiarity (activity, binarity, pollution by e.g. carbon) might not systematically affect the luminosity with respect to a star of similar nature but not peculiar.  This weak signature, however, might be significant enough to affect the pipeline deriving parameters such that the pipeline converges at wrong parameters (see further discussion in next section).   

If the signature is large, the stars normally will be more isolated in the \tsne\ map and not many normal stars will be in the same region (see also discussions in T17 on this matter), which can then affect the luminosity more significantly.  This was studied by \cite{2012MNRAS.425..355R}, who used the known and somewhat constant distance for all objects in the LMC to compare brightnesses of emission-line stars according to spectral types. In this case the constant was the distance rather than the spectral type.  They found that H$-\alpha$ alone could distort the expected magnitude according to spectral type by as much as 4 magnitudes. In RAVE we do not have H$-\alpha$ emission but it is certainly important to keep in mind that stars showing strong peculiar features might have uncertain distances with our method.  We find that stars flagged as hot (orange) show some systematically larger scatter in distances but still not significantly off from normal star scatter, in particular for the cool stars.  This is consistent with our results found for the clusters Blanco~1 and Melotte~22, whose members were hot stars (below).



\begin{figure*}
	\centering
	\includegraphics[width = \textwidth]{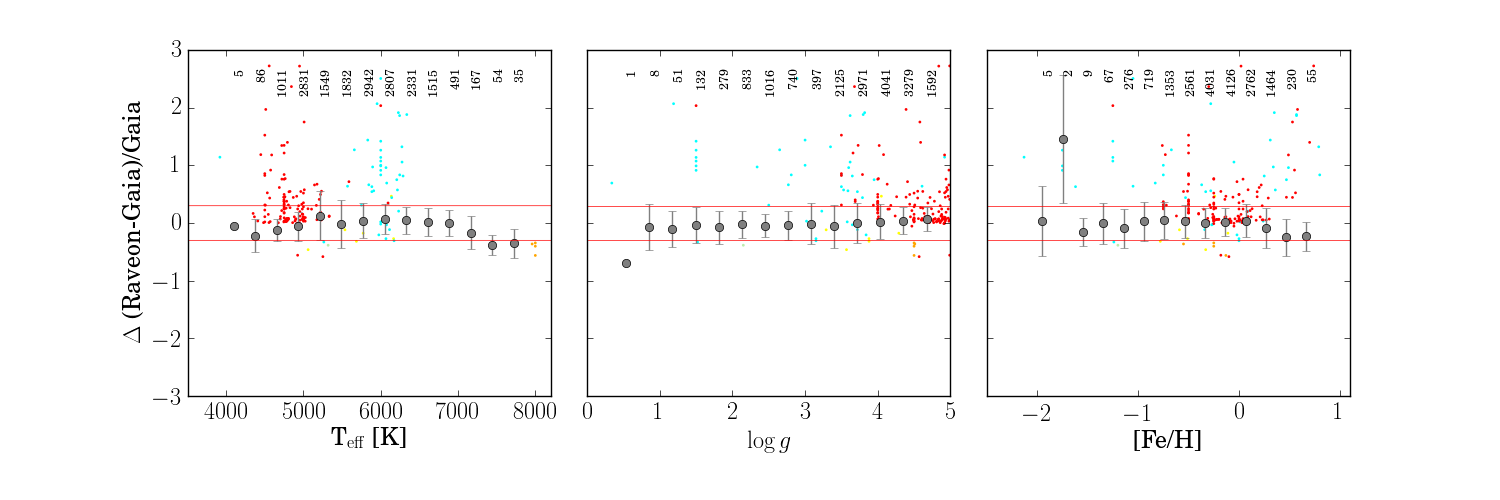}
	\includegraphics[width = \textwidth]{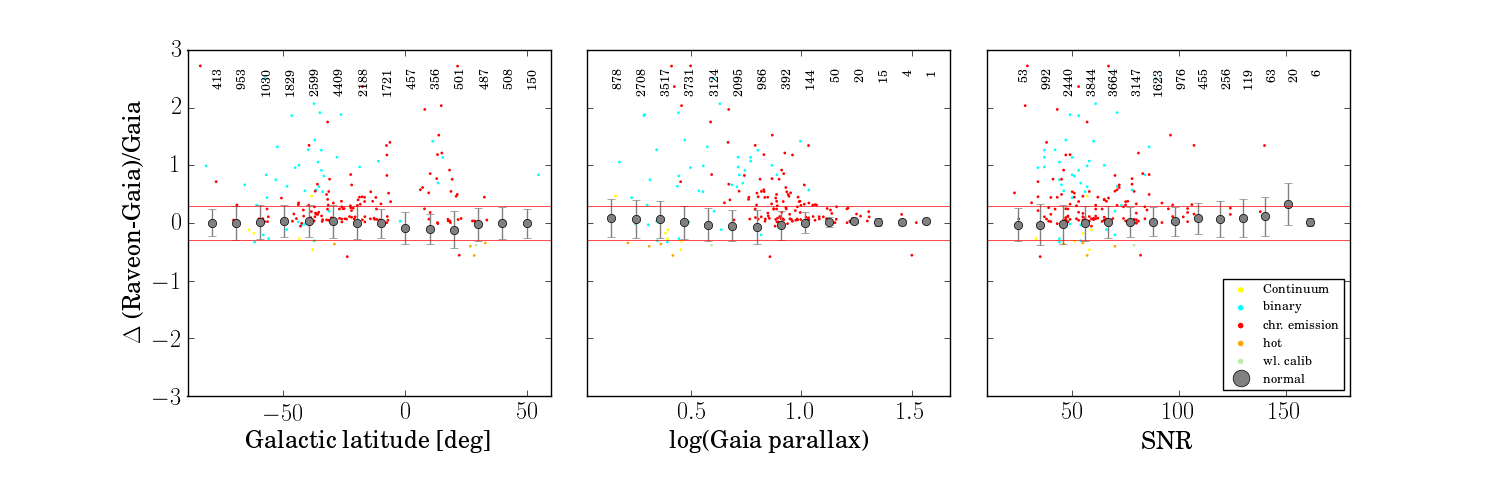}
    \caption{As in \fig{plx_params} but here using distances obtained from RAVE-on parameters and isochrones. }
    \label{plx_params_on}
\end{figure*}

\subsubsection{RAVE DR5 versus Gaia parallaxes}


Distances  and parallaxes to all RAVE stars  are provided in K17 as part of the RAVE-DR5 data release. They are obtained using the stellar parameters and with the help of PARSEC isochrones \citep{2012MNRAS.427..127B} as described in B14. We use these results in order to compare the performance of our results with an approach that is model-dependent.  Figure~\ref{plx_params_dr5} shows the same as  \fig{plx_params} but here the parallaxes are those provided by RAVE-DR5.  There is an overall similar picture than our case. 
Regarding the dependency of stellar parameters, the RAVE-DR5 parallaxes show a strong trend with effective temperature which is  an opposite one as for the twin parallaxes.  This reflects the fact that it is more difficult to determine distances for giants using isochrones given the large change in luminosity for small changes of temperature which characterises the red giant branch in the HR diagram \citep[see discussions in e.g.][]{2014MNRAS.445.2758R, 2016A&A...585A..42S}. 

It is important to note that parameters of red giants are also challenging to obtain. On the one hand, 3D effects become dominant in the atmospheres of giants \citep{2007A&A...469..687C, 2015A&A...573A..90M} and on the other hand, red giants have stronger lines making it more difficult to identify the continuum for spectral analysis \citep{2012A&A...547A.108L, 2014A&A...564A.133J, 2015A&A...582A..49H}. Differences of stellar parameters, in particular surface gravities, can yield  dramatic differences in distances of giants as extensively discussed in \citet{2013MNRAS.429.3645S}.    In B14 it is extensively discussed how the distances to giants are more uncertain than the distances to dwarfs.  As in our case, metal poor stars show a systematic offset but we repeat that there are too few metal-poor stars to have strong conclusions on this offset. 

When looking at $\Delta$ as a function of Galactic latitude, we again do not find a strong trend, but at lower latitudes ($15^{\circ} < b < 15^{\circ}$) an underdetermination of parallaxes (i.e. overestimation of distances) is obtained by DR5 with respect to Gaia. That again hints towards some uncertainties related to the treatment of extinction, but again this underdetermination is small compared to the uncertainties obtained for the giants. In this case, while the dispersion of $\Delta$ obtained for low parallaxes is larger than for high parallaxes, the systematic offset is not as large as in the twin case. Finally, the dispersion  of $\Delta$ seems to slightly improve for higher SNR stars, reflecting the better quality of the stellar parameters and hence the distances derived from them.



In contrast to our results, spectroscopic parallaxes obtained with DR5 parameters show larger discrepancies with respect to Gaia for binary stars (cyan), as well as for spectra with chromospheric activity (red).  It is expected that flagged stars will tend to yield more discrepant parallaxes with respect to Gaia since features in the spectra like extra lines due to a binary or emission due to chromospheric activity can lead to wrong solutions in the parameter determination pipeline.   If the pipeline to derive stellar parameters fails for these cases,  this can lead to wrong distances as we see here.  As in the previous case, hot stars are slightly more scattered than normal stars. We emphasise here that in K17 and in \cite{2013AJ....146..134K} it is extensively discussed how the parameters obtained for these flagged stars should not be used.


\subsubsection{RAVE-on versus Gaia parallaxes}

Parallaxes can be determined following the procedure described in B14 but considering the RAVE-on parameters instead. The comparison of these values  can be seen in \fig{plx_params_on} displayed in the same way as in \fig{plx_params} and \ref{plx_params_dr5}.  

In general the scatter between these parallaxes and those of Gaia is comparable to the ones obtained considering DR5 parameters or the twin method(see above) but with some minor differences for particular kind of stars, notably the systematic underestimation of Rave-on parallaxes for very hot stars. The hot stars have Rave-on parameters that are uncertain, because no suitable set of training set could be found \citep[see][for further discussions]{2016arXiv160902914C}.  The fact that the results are in general similar to the DR5 ones is not surprising since the stellar parameters of DR5 and RAVE-on are in general in good agreement when errors are taken into account (K17).  We comment that this method to determine distances is based on a Bayesian approach which uses the stellar parameters  and a Galactic model as a prior. If the parameters agree within errors, it is expected that the posterior probability for the parallax will be very similar in both cases. Besides the stellar parameters, both approaches employ the same data (photometric bands and isochrone sets) and further priors such as the Galaxy model and extinction law.  Indeed, the same systematically lower parallaxes are obtained for stars in the Galactic disk as in the previous cases.  For low metallicity, however, a similar large discrepancy of parallaxes as in the twin case is obtained, although this result is found for a very small sample of stars so the offsets could be caused by the small number  statistics. There is no significant trend of dependency on SNR or Galactic latitude for this case. 


The fractional difference between normal and flagged stars is quite different. While the normal stars appear to yield good agreements within 30\%, many chromospherically active stars and binaries do not, in a even more evident way than the DR5 or the twin case. This does not pose a serious problem for distances obtained with this method as stars are flagged and thus can easily be removed from the analysis. However, the difference between normal and flagged stars in this case suggests that the training set needed by {\it The Cannon} might be incomplete.  While recent results of parameters obtained with this method have shown an impressive performance \citep[e.g.][]{2015ApJ...808...16N, 2016ApJ...833..262H, 2017ApJ...836....5H} it is clear that this kind of methodology can only perform as well as the training set used to build the model.  It is still challenging to have a training set which contains representatives of all kinds of stars in a dataset, and that this training set has well-determined parameters for all of these stars. Since {\it The Cannon} uses the entire spectrum to derive a large set of labels, the fact that flagged stars show a large scatter in the derived parallaxes hints at this method employing the wrong model for the stars with peculiarities  \citep[see also][for further discussions]{2016arXiv160902914C}.





\begin{figure*}
	\centering
	\includegraphics[scale=0.25]{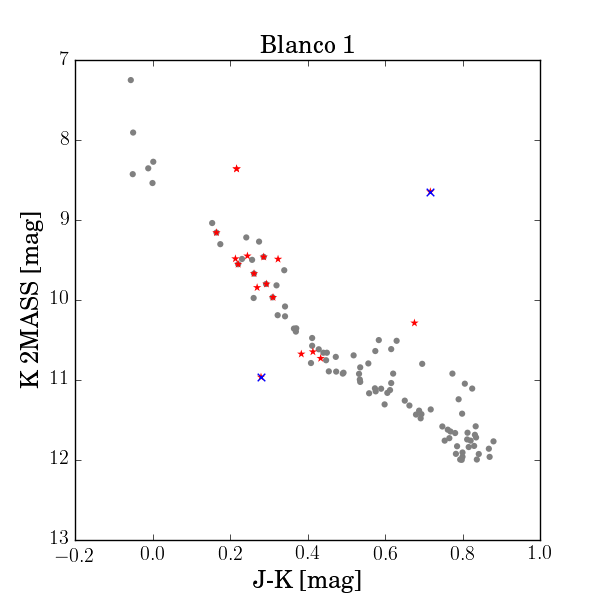}
	\includegraphics[scale=0.25]{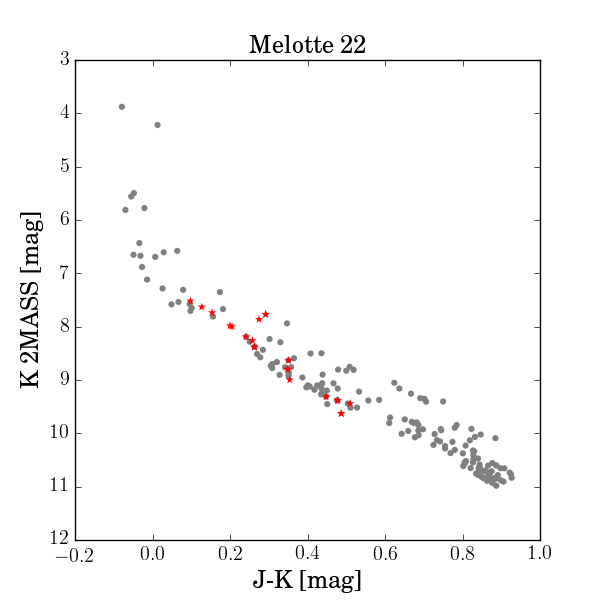}
	\includegraphics[scale=0.25]{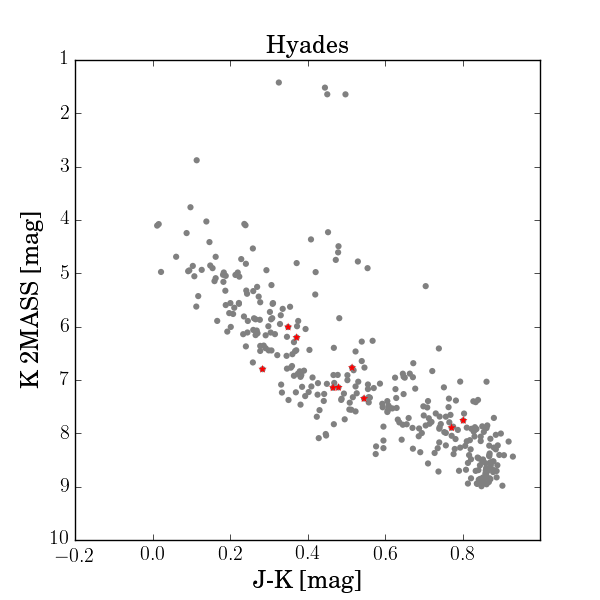}
	\includegraphics[scale=0.25]{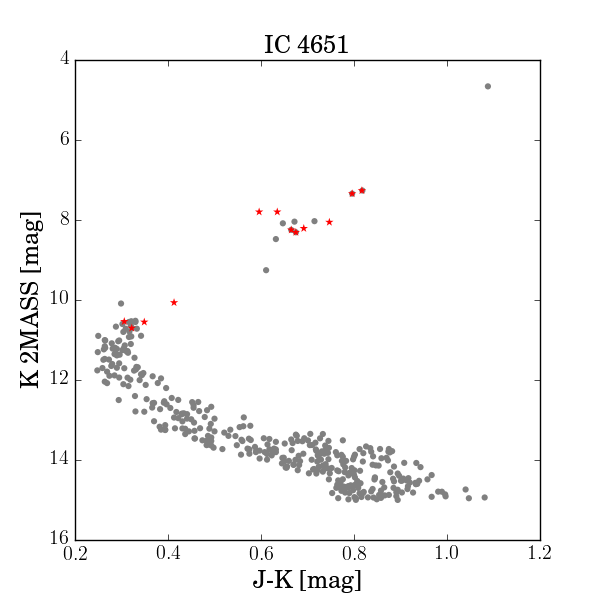}
	\includegraphics[scale=0.25]{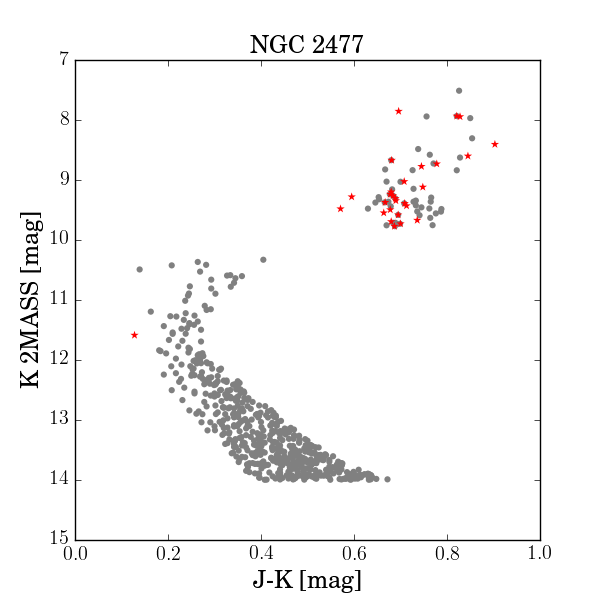}
	\includegraphics[scale=0.25]{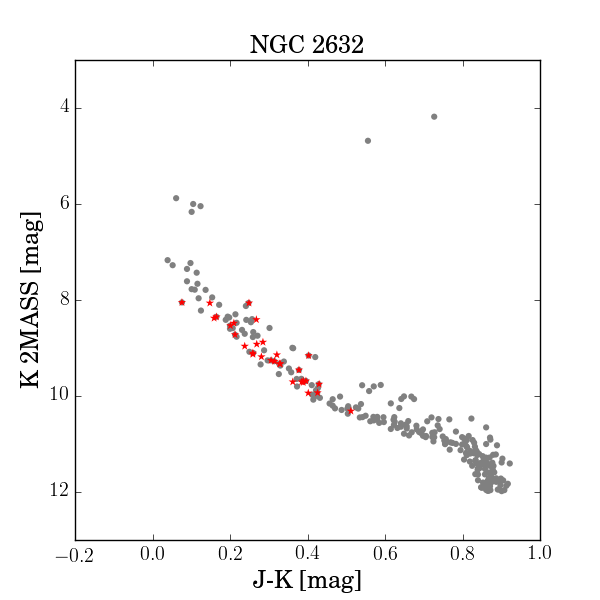}
	\includegraphics[scale=0.25]{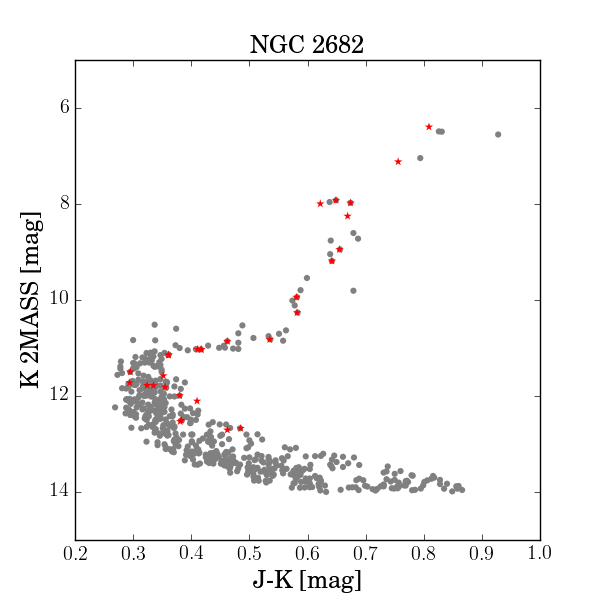}
    \caption{Open cluster 2MASS color magnitude diagram. In grey the stars from \citet{2013AA...558A..53K} which are selected to be members are shown while in red the RAVE stars that satisfy our selection criteria (see text) are plotted. }
    \label{cmds}
\end{figure*}

\subsubsection{Summary of comparisons}

We conclude that our parallaxes are competitive with those obtained from isochrones and stellar parameters. We emphasise that our parallaxes are model free. Furthermore, we can obtain parallaxes with equal confidence for the peculiar stars, except some with bad continuum normalisation. A better reduction of the data would be necessary to improve these distances. In the same way, better projection parameters for the \tsne\ map and perhaps further considerations of colours might help to separate better dwarfs and giants and thus decrease the scatter down to the level of the model-dependent distances. Such improvements remain as part of future work.  
  We note however that our results use the Gaia parallaxes of the remaining 75\% of the TGAS-RAVE sample and therefore they are calibrated on those data. The spectrophotometric parallaxes, on the other hand, are independent of astrometry, still making them competitive for testing Gaia parallaxes.

\subsection{Open clusters}\label{clusters}

Members of several  open clusters have been observed in RAVE. A list is reported in \cite{2014A&A...562A..54C},  providing an average value for their radial velocities and metallicities  as determined from RAVE spectra.  We looked for these members by selecting stars within an area of 2~degrees around their coordinates. We further selected the stars in that area that have RAVE-DR5 metallicities and radial velocities within 0.5~dex and 3~km/s, respectively, of the reported values in \cite{2014A&A...562A..54C}.  For our analysis we considered only the clusters for which at least 10 candidate members passed these selection criteria. For Melotte~22, we adopted the RV reported in \cite{2002AA...389..871D}, which gave us more selected members. We note that this cluster contains stars that are fast rotators, making the RVs and parameters of RAVE reported in  \cite{2014A&A...562A..54C} more uncertain.  We further included the Hyades cluster which was not studied in  \cite{2014A&A...562A..54C} but observed with RAVE. For that cluster we took the reported RV from the Simbad database \citep{2000A&AS..143....9W} and the metallicity from \cite{2014A&A...561A..93H}, which corresponds to a compilation of many literature reports.   In \tab{cluster_fin} we list the literature parallax, which is the mean and the standard deviation of different values reported by independent works (see \tab{cl_dist_lit}), as well as the reference RV and [Fe/H] considered for the clusters. We further list  our final result for the  twin parallaxes.

\begin{table}
	\centering
	\caption{Clusters analysed in this work. The reference parallax ($\varpi_L$) the metallicity ([Fe/H]) and radial velocity (RV) are obtained from the literature (see text). Our results for the parallax are listed as $\varpi_{Tw}$ with N representing the number of members used for the analysis.  }
	\label{cluster_fin}
	{\small
	\begin{tabular}{|l|r|r|r|r|r|r|r|}
\hline
  \multicolumn{1}{|c|}{Cluster} &
  \multicolumn{1}{c|}{$\varpi_L$} &
  \multicolumn{1}{c|}{$\sigma \varpi_L$} &
  \multicolumn{1}{c|}{[Fe/H]} &
  \multicolumn{1}{c|}{RV} &
  \multicolumn{1}{c|}{$\varpi_{Tw}$} &
  \multicolumn{1}{c|}{$\sigma \varpi_{Tw}$ }& 
  \multicolumn{1}{c|}{N}\\
& [mas] & [mas] & [dex] & [km/s] & [mas] & [mas] & \\
\hline
  NGC 2477 & 0.77 & 0.03 & -0.19 & 6.37 & 1.01 & 0.11 & 14\\
  Blanco 1 & 4.16 & 0.14 & -0.19 & 6.17 & 4.03 & 0.57 & 7\\
  Melotte 22 & 7.38 & 0.31 & -0.04 & 5.9 & 6.34 & 0.21& 8\\
  NGC 2682 & 1.18 & 0.04 & -0.1 & 33.81 & 1.43 & 0.11& 14\\
  NGC 2632 & 5.4 & 0.06 & 0.1 & 34.0 & 5.54 & 0.28&19\\
  Hyades & 21.29 & 0.81 & 0.0 & 39.4 & 23.70 & 1.46&6\\
  IC 4651 & 1.09 & 0.04 & -0.13 & -30.41 & 1.59 & 0.15&8\\
\hline\end{tabular}

	}
	\end{table}
 
 \begin{figure}
	\centering
	\includegraphics[width = \columnwidth]{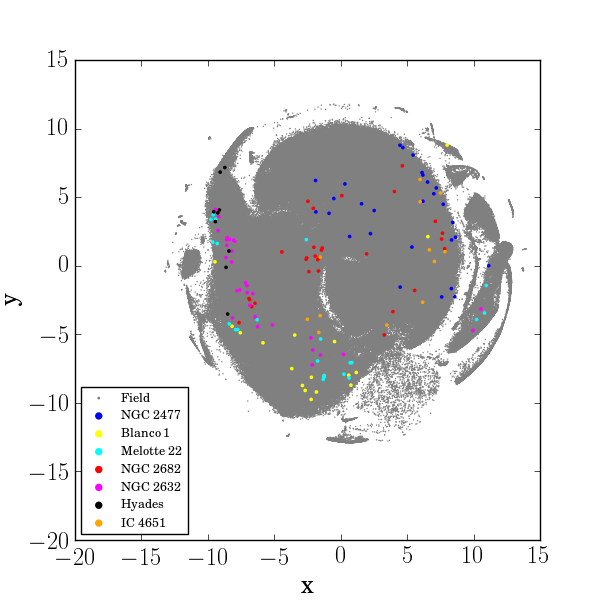}
    \caption{\tsne\ map  with the cluster members indicated with colours. We have cluster stars spanning a large fraction of the parameter space of RAVE. }
    \label{clusters_tsne}
\end{figure}

\begin{figure}
	\centering
	\includegraphics[scale = 0.45]{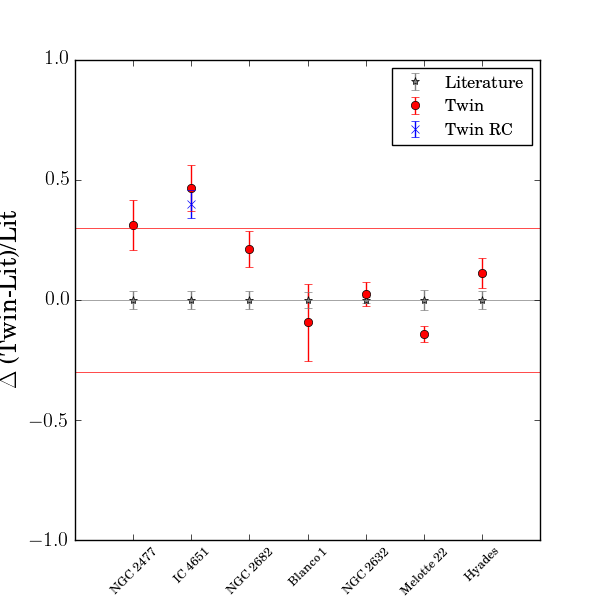}
    \caption{Difference between our values compared with the mean literature value for each cluster as indicated in \tab{cluster_fin}. The clusters are sorted by parallax, i.e. NGC 2477 is the most distant cluster and the Hyades the closest. The red lines indicate the 30\% agreement. Most of the clusters, when considering the errors, agree within 30\% with the literature.}
    \label{cluster_results}
\end{figure}

A further check to ensure we had selected the right cluster members was to analyse the colour-magnitude diagrams (CMD) of the clusters using 2MASS photometry. We considered the stars analysed in \cite{2013AA...558A..53K}, selecting the stars which had a spatial probability $P_s=1$, kinematic probability $\mathrm{Pkin}>0.68$ and photometric probability $\mathrm{P_{JK}}>0.68$.  This was done for all clusters except the Hyades, for which we took the photometry published in  \cite{2011A&A...531A..92R}.  In the latter, no further selection based on membership probability was applied since the photometry published in that paper had already analysed  membership probabilities.  

 The CMDs of the selected stars can be seen in \fig{cmds} with grey. Our RAVE targets are overplotted with red star symbols. There are two clear outliers in Blanco~1 which have a blue cross and have been removed from our analysis, but for the rest, the CMDs tell us that our selection of members is robust. In \fig{clusters_tsne} we plot the \tsne\ map but this time the cluster stars are colour-coded by cluster membership. The grey colour indicates the field stars. We can see that we have cluster members that span a large area of the map, meaning that we are not biasing this analysis towards one specific spectral type only.

Our results were obtained  from the average of the individual twin parallaxes determined for each of the cluster members  indicated with red stars in the CMDs of \fig{cmds} with the uncertainty corresponding to the standard error of the mean. The names of these member stars along with the twin parallaxes can be found in \tab{individual_values}. A comparison of the parallaxes for all clusters is shown in \fig{cluster_results}, where we plot the fractional difference, $(\varpi_{\mathrm{Twin}} - \varpi_{\mathrm{Lit}})/\varpi_{\mathrm{Lit}}$, for each cluster sorted by parallax, i.e, from most distant to closest cluster.   Our results are given in red while in grey the error bars of the literature are shown. We see a general good agreement when considering the error bars, with no strong dependence on cluster distance.  

The largest difference is found for IC~4651.  However, if we consider only the red clump stars of the cluster, our resulting {  parallax of $1.4\pm 0.06$~mas agrees} better with the literature value. That value is indicated with a blue cross in \fig{cluster_results}. The dwarf stars in the clusters are hot and are located in a region of the \tsne\ map in which degeneracies with cool giants exist \citep[see \fig{rave_tsne} and \fig{clusters_tsne}, and also][]{2011A&A...535A.106K}.  It is possible that for these stars many giants were used as reference, underestimating the resulting distance. 

We comment that for NGC 2682 (also known as M67) and for Melotte~22,  a twin parallax has already been reported using high-resolution spectroscopy from HARPS in J15 ($\varpi = 1.23$ mas for NGC~2682) and in M16 ($\varpi = 7.45$~mas for Melotte~22). They present an opportunity to compare the performance of the twin method between high resolution and extended wavelength coverage with lower and shorter wavelength coverage. Our results agree within 20\%. Melotte~22  is a young cluster and the cluster members observed with RAVE are many fast rotators and very hot, which means that the spectra do not contain as many signatures as older and of later spectral types would, e.g., the cluster members we found for NGC 2682 (see \fig{cmds}).  Blanco~1 is a cluster with similar age and metallicity properties as Melotte~22, and thus we obtain a similar overestimate of the distance. The fact that in M16 the distance of the cluster is more accurately determined indicates that the twin method can be applied for such stars, but that the spectra need to be of higher resolution. 

Taking our results without further cuts on spectral type, we obtain a { mean agreement of  $0.13 \pm 0.19$,} which is consistent with the external uncertainties of our method (see \sect{uncertainties}).

\begin{figure*}
	\centering
	\includegraphics[width = \columnwidth]{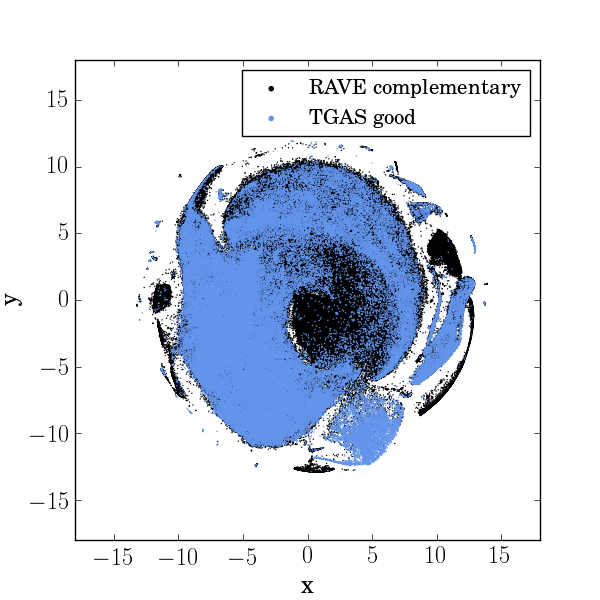}
	\includegraphics[width = \columnwidth]{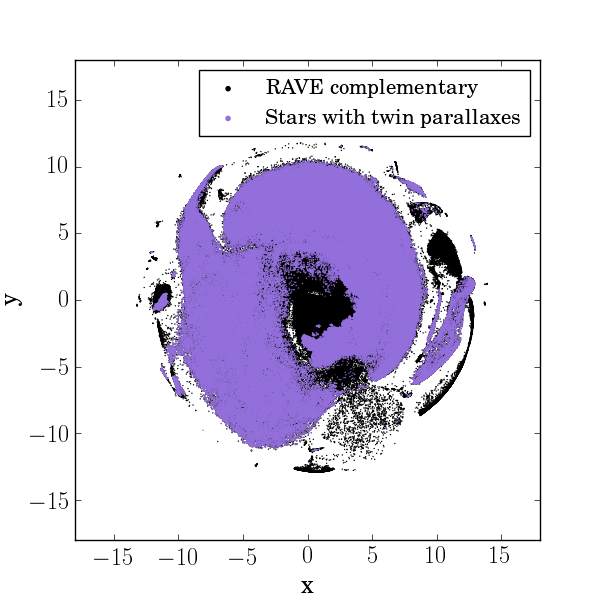}
    \caption{\tsne\ maps. Left panel:  the complementary RAVE dataset is plotted with black and the RAVE-TGAS stars with positive and accurate parallaxes are displayed with blue. Right panel: The complementary RAVE dataset is plotted with black and the stars for which we could determine distances  using the  RAVE-TGAS stars with positive and accurate parallaxes as reference are displayed   with magenta. }
    \label{twins_tsne}
\end{figure*}

\begin{figure*}
	\centering
	\includegraphics[width = \textwidth]{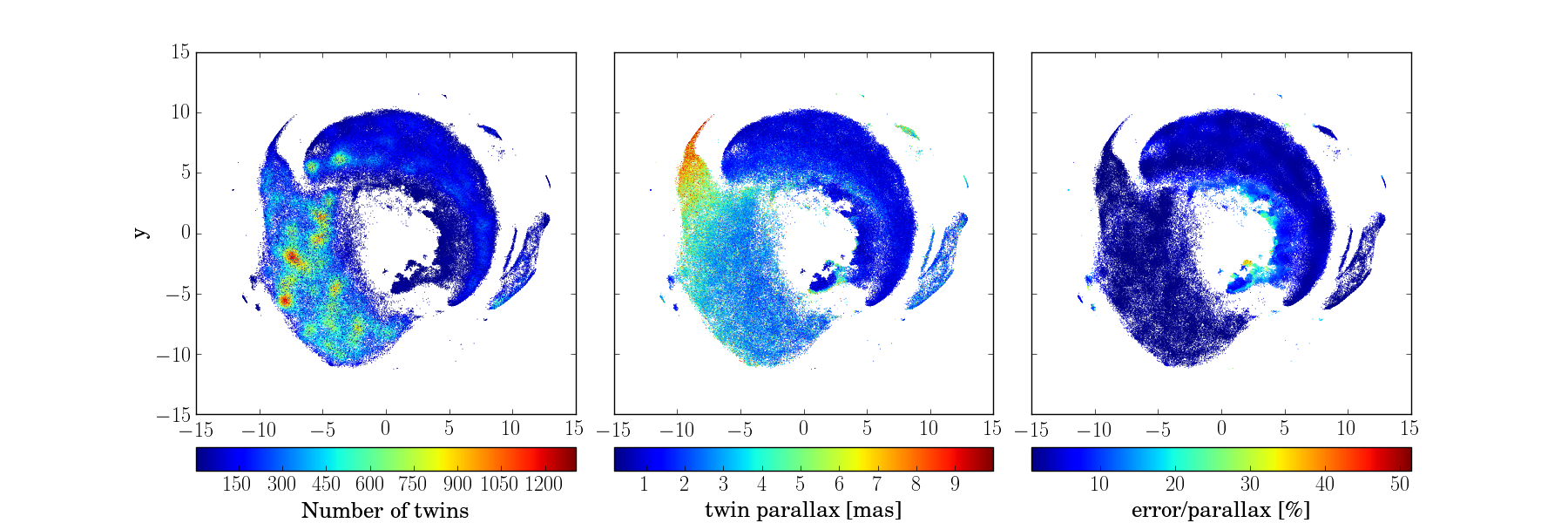}
    \caption{\tsne\ map of the stars with twin parallaxes coloured by number of twins (left), parallax (middle) and fractional error of the parallax (right). }
    \label{plx_tsne}
\end{figure*}

\section{Discussion}\label{catalogue}

\subsection{Parallaxes of the complementary RAVE sample}

The complementary RAVE sample consists of stars with negative TGAS parallaxes or with TGAS parallaxes with uncertainties above 20\%, as well as the RAVE stars which are not part of TGAS. This sample contains 367,895 stars with non-saturated 2MASS photometry ($K_s > 4.5$ mag) and spectra with $\mathrm{SNR} > 10$. 

In the left panel of  \fig{twins_tsne} we show the \tsne\ map these stars with black. We also show the \tsne\ map of the TGAS-RAVE sample with blue selected to have non-saturated 2MASS photometry, spectra with SNR above 1, positive parallaxes and parallaxes measured with accuracies better than 20\%. This is our sample used to find twins for reference parallaxes. We can see that there is a good overlap of both samples in the 2D \tsne\ map. This indicates that we have a good chance of finding twins for most of the complementary RAVE sample and hence determining parallaxes.  We however note the clear islands for which there are no reference TGAS stars, in particular in the regions of cool giants at the centre, as well as the edges of the map. 
 
   By requiring $R<0.5$, $\chi^2 \leq670$, and at least 5 Gaia targets in the TGAS-RAVE catalogue (see \sect{twinicity}), we could determine parallaxes for  { 232,545}  targets. This corresponds to 60\% of the entire complementary RAVE sample. This is less than what we found in \sect{twinicity} but we were biased towards the same type of stars while in \fig{twins_tsne} we see that the TGAS targets do not cover the entire domain of the complementary RAVE sample.   
   
   In the right panel of \fig{twins_tsne} we plot again the projections of the entire complementary RAVE sample with black but we overplot the stars for which we could determine distances with magenta.  We find twins for essentially all stars in the central parts of the map, which are almost all normal stars according to the flags of  M12 (see also \fig{flags_tsne}). 
  We also find twins for stars located some of the islands of the map, notably for binaries, and chromospherically active stars.  As  discussed extensively above, our method does not show significant discrepancies in the parallaxes retrieved for these peculiar stars with respect to normal stars and therefore the parallaxes derived with our method are as reliable as for normal stars. We recall  however that stars flagged with problems with continuum could be incorrect, as well as stars flagged as very hot.  
   

Figure~\ref{plx_tsne} shows again the \tsne\ map this time coloured by the number of TGAS twins found for each target (left), the parallax (middle) and the fractional internal error of the parallax (right).  There are few spots in the \tsne\ map which are very dense where we find more than 1000 twins, but the majority have about 300 twins.  In addition, there are some other regions in which fewer than 50 twins are found. 

 The middle panel of \fig{plx_tsne} is interesting as it shows how the giants are more distant than the dwarfs. We know from \fig{rave_tsne} that the left  region of the \tsne\ map  is where the FGK dwarfs lie while the upper right hand side is where the FGK giants are located. It is consistent that the map colour distribution has a tendency to be bimodal. The fact that there are very few nearby giants hints at selection effects of RAVE observing all stars within a magnitude range. Nearby giants are too bright and therefore are either not part of RAVE or have saturated photometry which implies we cannot use Eq.\ref{eq1} for the parallax determination \citep[see ][for details of the selection function]{2016arXiv161100733W}.     
 
 We further note that the closest stars correspond to the cool dwarfs (see \fig{rave_tsne}). They populate  the upper tip of the map, as well as a small region above the FGK giants around $(x, y) = (6,6)$.  They have to be close since their intrinsic luminosities are very low.  The island of hot stars also corresponds to mainly OBA main-sequence stars, which are intrinsically brighter than FGK dwarfs but fainter than giants. These stars have a wide range of parallaxes.    There are other islands around  $(x, y) = (-10, 0)$ which contain rather distant stars. Most of these islands are where the very cool giants lie (see \fig{rave_tsne}). These stars are very luminous so it is expected they are in general more distant.  

It is also interesting to see a rather sharp transition from nearby to distant stars at the centre of the map. The parameters also change relatively abruptly in that region of the map. There is a mixture however of dwarfs and giants, which explains why in that region the error in the parallax is larger (see right panel).  When inspecting that panel, we also note that the fractional internal error in the parallax is in general independent on the parallaxes.  Other more uncertain results are seen for the island at  the inner edge of the right hand side of the map. These regions are flagged to have problems with the continuum, have very few twins (see right hand panel) and correspond to the regions in which dwarfs and giants overlap, causing a larger dispersion of the distribution of twin parallaxes. The stream of the binaries does not apear to be significantly more uncertain than the rest of the stars. 

We note, however, that the island of cool giants has quite uncertain parallaxes. These stars are not flagged, but we know they show evidence of being discrepant from TGAS in our analysis of \sect{comparisons}. These  stars also have a small number of twins, making the standard error of the mean systematically larger than the rest of the sample.  To summarise, we obtain in general internal errors that are no more than 10\% the majority are of the order of 5\% but in a few cases of the order of 30\%. 

\begin{figure}
	\centering
	\includegraphics[width = \columnwidth]{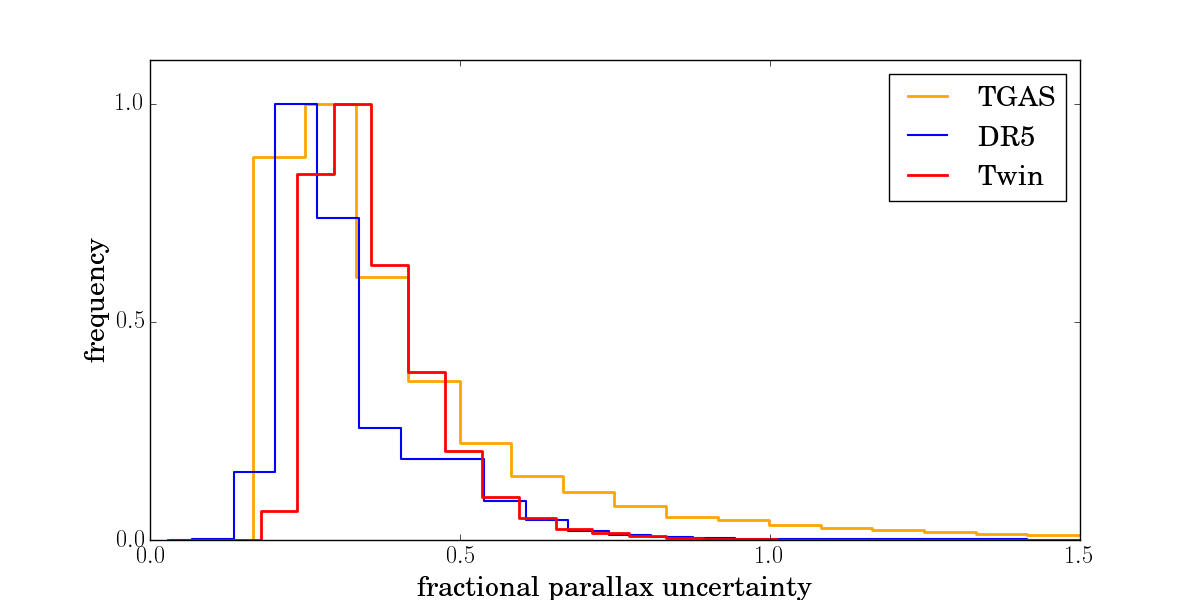}
    \caption{Distribution of fractional uncertainties for the TGAS stars which were not included as reference stars, along with the results obtained with the twin method, as well as with DR5. }
    \label{histo_errors}
\end{figure}

\subsection{Comparison with TGAS}

In the complementary RAVE sample there are 36\% of the stars that have positive TGAS parallaxes with accuracies worse than 20\%. We compare these values with those obtained with the twin method, as well as with the spectroscopic parallaxes from DR5. In our case we consider the total uncertainties, i.e.  the internal standard error of the mean and the external one of 28\% obtained from our analysis in \sect{uncertainties}, which are added quadratically.  

In \fig{histo_errors} we show the distributions of the total fractional uncertainties obtained by us with the red histogram while the TGAS uncertainties are shown with orange and DR5 with blue. All histograms have been scaled to peak at 1.0 for better visualisation. The quality cuts on the parallax of requiring accuracies better than 20\% were done on the reference sample for our results. Our results peak at higher uncertainties than TGAS and RAVE-DR5, but TGAS has a long tail towards uncertainties above 50\%.  The results from RAVE seem in general to be more accurate than ours, but since they depend on stellar parameters, which means that for 5\% of this sample no DR5 parallaxes were available. 

This comparison shows us that the twin method is a good complement of Gaia as it helps to improve the precision for uncertain cases, even when spectra of short wavelength coverage and intermediate resolution are to our disposal. This situation can only improve in the advent of  on-going and future high resolution spectroscopic surveys.

 \subsection{Climbing the cosmic distance ladder in RAVE}
 
\begin{figure}
	\centering
	\includegraphics[width = \columnwidth]{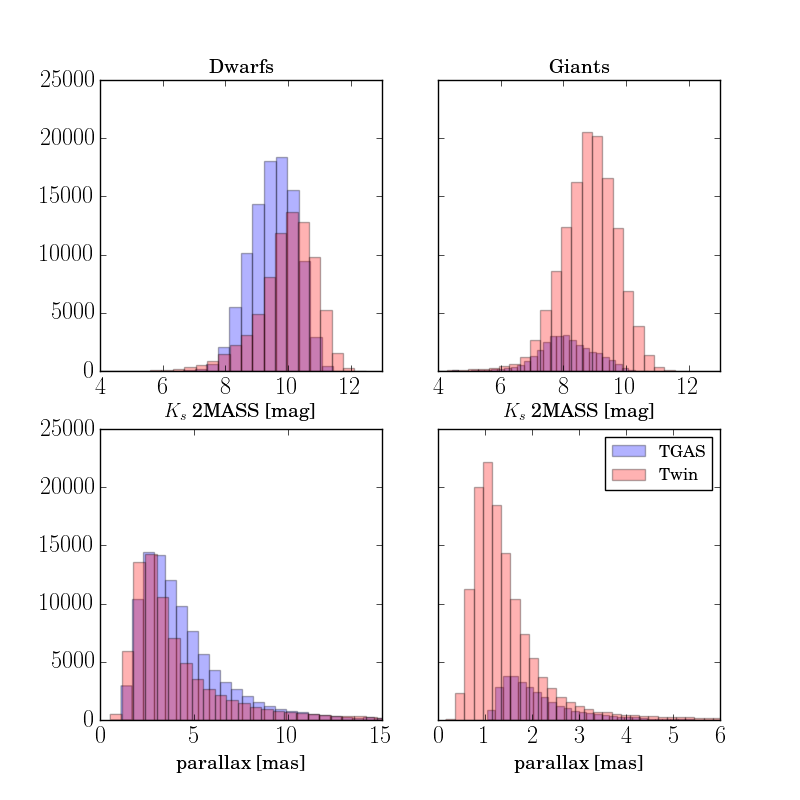}
    \caption{Distribution of $K_s$ magnitudes  (top) and parallaxes (bottom) for dwarfs (left) and giants (right) of the TGAS and our complementary sample, in blue and red, respectively.  }
    \label{histo_mags}
\end{figure}

 Due to the large differences in absolute magnitudes between dwarfs and giants, we separated the sample using the colour-magnitude diagram constructed using the Gaia parallaxes for the TGAS-RAVE sample with accuracies above 20\% (reference TGAS sample) and our twin parallaxes for the complementary RAVE sample.  The $K_s$ and parallax distributions both samples for dwarfs and giants are shown in \fig{histo_mags}. 
 
In the upper panels of \fig{histo_mags} we plot the $K_s$ distributions of dwarfs and giants separately for both, reference TGAS and the complementary RAVE samples, with blue and red colours, respectively.  The complementary RAVE sample is in general fainter than the TGAS one but there is still a large overlap in brightness between both samples.  The giant sample in particular shows a very large number of complementary RAVE stars that are fainter than the TGAS reference sample.  

The distribution of parallaxes is shown in the lower panels of \fig{histo_mags}, following the same nomenclature and order as the upper panels.  For the dwarfs, it seems that the twin stars cover the same range in distances our reference TGAS sample,  which might suggest that the faint end of the $K_s$ distribution of the dwarfs corresponds mostly to faint low-mass dwarfs which are not very luminous.  For the giant sample, however, the twin stars are overall at large distances compared to the  reference TGAS ones.

\subsection{Final catalogues}
We provide a catalogue of twin distances to 232,545 stars with internal uncertainties defined as the standard error of the mean of all  individual twin parallaxes. They can be downloaded from RAVE page or from Vizier. 

We also provide the list of twins candidates for each of the TGAS stars  used in our reference sample as selected from the \tsne\ maps and their value for the $\chi^2$. This material can be used to develop further models of parallax estimate of twin stars selected from our selection of candidates, as well as to investigate further science problems of galactic structure of twin stars.

\section{Summary and Conclusions}\label{conclusions}

We have described the application of the twin method to determine parallaxes for stars in the RAVE survey.  The method is based on the proposition that if two stars have the same spectra (hence twins) they have the same luminosity. By knowing the parallax of one star we can estimate the parallax of its twin by comparing their magnitudes. The power of this method is that it is remarkably simple and model-independent. 

To  find twin stars,  we applied the t-Student stochastic neighbour embedding technique (\tsne) to analyse the spectra in a two-dimensional map in which the spectra are distributed according to their similarity. Neighbouring stars in the map are treated as twin candidates.  The reason for using \tsne\ to do this was because the RAVE dataset is the largest one used with the twin method so far, and the spectra have more noise and shorter wavelength coverage than previous works on this subject. We therefore had to implement a procedure to deal with the entire available spectrum efficiently. Previous literature on the performance of \tsne\ and related techniques on spectra suggested that this procedure is suitable for our purpose.  In this paper we demonstrate that it is the case. We further computed for all the twin candidates their $\chi^2$ to select the best twin candidates. 

By requiring the stars located at a radius of $R=0.5$ in the \tsne\ map around a given target to be their twins, and selecting those with $\chi^2 \leq 670$ we were able to determine parallaxes for 60\% of the sample. This number was obtained after selecting only the stars with TGAS parallaxes more accurate than 20\%, which reduced our reference sample by a factor of two and removed a large portion of the cool giants.  The accuracies of our parallaxes were investigated and compared to the accuracies of other model-dependent methods based on stellar parameters derived from the spectra and isochrone fitting. Our uncertainties show a unit dispersion  when compared to Gaia parallaxes, which suggests that our results have properly estimated systematic uncertainties. We defined the internal uncertainties as the standard error of the mean of all twins found in the map with the criterion of above. External uncertainties were  estimated as 28\% from comparisons to TGAS. We applied the method to 7 open clusters, finding  a similar agreement with the literature as with our comparisons with TGAS.    

An important result of this work is that we could also determine  parallaxes with comparable accuracy for peculiar  and for normal stars. This result further demonstrates the power of the twin method in that it does not depend on stellar parameters. Peculiar signatures in spectra are sometimes very weak and do not affect significantly the total stellar  flux, hence maintaining the proposition that they have same luminosities. These weak signatures can however cause a significant effect in parameter determination leading to erroneous solutions. This propagates to wrong distances with model-dependent methods. Likewise, cool dwarfs have isochrones that are uncertain, also making it more difficult to determine distances for these stars. Our twin parallaxes for them are like for any other star as we do not employ isochrones. 

We note however that for stars with low gravities our results were not very accurate, mainly because of degeneracies between dwarfs and giants in RAVE spectra which means that we had dwarf stars as reference for giants.  Very hot stars also showed more uncertain results,  perhaps due to the few and very broad lines presented in the spectra which makes the radial velocity determination and thus the comparison of the spectra harder.  Finally, we found that for cool giants our results also showed larger errors. We attributed that to the fact that for those stars  fewer twins were found leading to a larger standard error of the mean.

Our final catalogues include a list of all twin candidates for each TGAS-RAVE target of our reference sample, as well as a catalogue of twin parallaxes of  the complementary RAVE targets. These catalogues can be found either on the RAVE website\footnote{\url{https://www.rave-survey.org/project/}} or in Vizier. \\

\noindent We conclude this work by answering the questions placed in the introduction:
\begin{enumerate}
\item {\it How accurate are our results  when applied to the RAVE survey?}  We can compare directly the results obtained with the twin method with very high resolution  and extended wavelength coverage from spectra taken from the HARPS spectrograph and with lower resolution and shorter wavelength coverage from RAVE spectra for the distances of two open clusters, the Pleiades (Melotte 22) and M67 (NGC 2682).  For the Pleiades we obtain with RAVE spectra an accuracy of 0.23~mas while  a much better accuracy of 0.03~mas was obtained with HARPS spectra. For M67 the results have comparable accuracies which might be due to the fact that the stars analysed in M67 are in the red clump, whose spectra have well-defined lines.  It is clear that the method applied to spectra like RAVE will produce more uncertain results, as the confidence with which we assess twin stars decreases, in particular for spectra which have very few and broad lines.

 While it is true that one pair of twins might yield a twin parallax that is significantly more uncertain in the RAVE case than in HARPS, the power of surveys is that we have much larger datasets at our disposal than the HARPS archive. Thus, the final results will have a strong statistical significance. In several cases, we found more 500 twins for our targets, which means that the mean of the distribution of their distances is very robust, with standard errors that tend to be negligible compared to the input uncertainty of Gaia. 

\item {\it For how many stars in a spectroscopic survey can we apply the twin method to determine distances? }  For RAVE, we could determine distances to 60\% of the sample. This is because the TGAS reference and the complementary RAVE samples have stars of very similar nature, which means that we could find reference stars for almost every target in the \tsne\ map which was covered by the accurate TGAS reference sample.  Spectroscopic surveys follow a selection function and thus obtain spectra that are similar to each other.  In the case of RAVE, we refer the reader to \cite{2016arXiv161100733W} for its selection function.  We also note here that the reference dataset is equally large as the target dataset.  It remains to be investigated if this percentage is also obtained for other surveys with a smaller overlap with TGAS, such as APOGEE. 
\end{enumerate}

\noindent Soon the second Gaia data release will become available and will contain parallaxes of a billion of stars in the Galaxy, including all stars of the RAVE survey. While such trigonometric parallaxes will provide with the basic knowledge about the distances of stars, the accuracy will inevitably decrease with distance.  With the datasets of stellar spectra  forthcoming with WEAVE \citep{2016SPIE.9908E..1GD} and 4MOST \citep{2012SPIE.8446E..0TD}, the twin method offers an attractive alternative to complement Gaia for the most distant stars. 

\section*{Acknowledgements}

This work was partly supported by the European Union FP7 programme through ERC grant number 320360. PJ acknowledges King's College Cambridge Research Associate Programme and C. Rodrigo for tips in using VO tools for this work. KH is supported by Simons Foundation.
This research has made use of the SIMBAD database,
operated at CDS, Strasbourg, France, as well as tools from the Virtual Observatory.  \\

This work has made use of data from the European Space Agency (ESA)
mission {\it Gaia} (\url{https://www.cosmos.esa.int/gaia}), processed by
the {\it Gaia} Data Processing and Analysis Consortium (DPAC,
\url{https://www.cosmos.esa.int/web/gaia/dpac/consortium}). Funding
for the DPAC has been provided by national institutions, in particular
the institutions participating in the {\it Gaia} Multilateral Agreement.\\

Funding for RAVE has been provided by: the Australian Astronomical Observatory; the Leibniz-Institut fuer Astrophysik Potsdam (AIP); the Australian National University; the Australian Research Council; the French National Research Agency; the German Research Foundation (SPP 1177 and SFB 881); the European Research Council (ERC-StG 240271 Galactica); the Istituto Nazionale di Astrofisica at Padova; The Johns Hopkins University; the National Science Foundation of the USA (AST-0908326); the W. M. Keck foundation; the Macquarie University; the Netherlands Research School for Astronomy; the Natural Sciences and Engineering Research Council of Canada; the Slovenian Research Agency; the Swiss National Science Foundation; the Science \& Technology Facilities Council of the UK; Opticon; Strasbourg Observatory; and the Universities of Groningen, Heidelberg and Sydney.
The RAVE web site is at https://www.rave-survey.org.

\bibliographystyle{mnras}
\bibliography{refs_rave_twin}

\begin{appendix}
\section{Open clusters}
In this section we list the individual values for the resulting parallax of each cluster member found in the RAVE survey in \tab{individual_values}.  The RAVE name of the star is given, as well as its corresponding cluster.  

\begin{table}
\centering
\caption{Parallaxes obtained for individual members in each cluster analysed in this work. A full version of this table can be found online.}
\label{individual_values}
\begin{tabular}{|l|l|r|r|r|}
\hline
  \multicolumn{1}{|c|}{cluster} &
  \multicolumn{1}{c|}{star} &
  \multicolumn{1}{c|}{$\varpi$ twin} &
  \multicolumn{1}{c|}{$\sigma \varpi$ twin} &
  \multicolumn{1}{c|}{ntw} \\
\hline
  Blanco 1 & 20050807\_0005m26\_005 & 5.79238 & 0.132197 & 428\\
  Blanco 1 & 20060726\_2353m28\_122 & 3.29293 & 0.0547224 & 673\\
  Blanco 1 & 20060930\_2359m33\_084 & 3.057 & 0.0594117 & 613\\
  Blanco 1 & 20061118\_2353m28\_117 & 4.22382 & 0.249239 & 84\\
  Blanco 1 & 20061118\_2353m28\_118 & 3.49395 & 0.111179 & 265\\
  Blanco 1 & 20110727\_2353m28\_116 & 5.37608 & 0.12647 & 467\\
  Blanco 1 & 20110905\_0020m31\_044 & 3.00227 & 0.10113 & 443\\
  Melotte 22 & 20120107\_0336p23\_107 & 5.6664 & 0.183508 & 252\\
  Melotte 22 & 20120107\_0336p23\_115 & 6.38786 & 0.208054 & 267\\
  Melotte 22 & 20120107\_0336p23\_119 & 6.35511 & 0.380769 & 39\\
  Melotte 22 & 20120109\_0346p23\_117 & 7.00413 & 0.282162 & 15\\
  Melotte 22 & 20120109\_0346p23\_127 & 5.68836 & 0.157212 & 366\\
  Melotte 22 & 20120109\_0346p23\_133 & 6.60906 & 0.334615 & 143\\
  Melotte 22 & 20120112\_0346p23\_117 & 6.44514 & 0.35437 & 17\\
  Melotte 22 & 20120112\_0346p23\_133 & 6.52762 & 0.348053 & 126\\
\hline\end{tabular}

\end{table}

Here we also list the literature values for the distance of the clusters in \tab{cl_dist_lit}. The parallax was obtained by inverting the distance as $1000/d$, with $d$ corresponding to the distance in parsecs, when only distance was provided. The reference for each value is listed too. In \tab{cluster_fin} we list the mean and the standard deviation of the values indicated here.

\begin{table}
\centering
\caption{Parallaxes reported in the literature for the clusters in our study. }
\label{cl_dist_lit}
\begin{tabular}{|l|r|l|}
\hline
  \multicolumn{1}{|c|}{cluster} &
  \multicolumn{1}{c|}{$\varpi$ [mas]} &
  \multicolumn{1}{c|}{ref} \\
\hline
  Blanco 1 & 4.00 & \citet{2002AA...389..871D}\\
  Blanco 1 & 4.34 & \cite{vL17}\\
  Blanco 1 & 4.14 & \citet{vL09}\\
  NGC 2477 & 0.74 & \citet{2002AA...389..871D}\\
  NGC 2477 & 0.80 & \citet{2004AA...423..189E}\\
  Melotte 22 & 7.41  & \cite{2002AA...389..871D}\\
  Melotte 22 & 7.69 & \cite{2013AA...558A..53K}\\
  Melotte 22 & 6.66 &\cite{2003AJ....125.1397C}\\
  Melotte 22 & 7.42 & M16\\
  Melotte 22 & 7.48 & \cite{vL17}\\
  Melotte 22 & 7.57 & \cite{2016AJ....151..112D}\\
  Melotte 22 & 7.44 & \cite{2017AA...598A..48G} \\
  NGC 2862 & 1.21 & \cite{2006MNRAS.371.1641P}\\
  NGC 2862 & 1.16 & \cite{1997AJ....114.2556T}\\
  NGC 2862 & 1.10 & \cite{2013AA...558A..53K}\\
  NGC 2862 & 1.22 & \cite{2016ApJ...832..133S}\\
  NGC 2862 & 1.23 & J15\\
  NGC 2862 & 1.14 & \cite{2009ApJ...698.1872S}\\
  NGC 2862 & 1.20 & \cite{2007ApJ...655..233A}\\
  Hyades & 22.22 & webda\\
  Hyades & 21.53 & \cite{vL09}\\
  Hyades & 21.39 & \cite{vL17}\\
  Hyades & 20.00 & \cite{1997AJ....114.2556T}\\
  NGC 2632 & 5.34 &\cite{2002AA...389..871D}\\
  NGC 2632 & 5.47 &\cite{vL17}\\
  IC 4651 & 1.13 &\cite{2013AA...558A..53K}\\
  IC 4651 & 1.05& \cite{1997AJ....114.2556T}\\
\hline\end{tabular}

\end{table}

\end{appendix}

\bsp	
\label{lastpage}
\end{document}